\documentclass[reprint,superscriptaddress,amsmath,amssymb, aps, prl, floatfix,]{revtex4-2}

\usepackage{graphicx} 
\usepackage{amsmath,amsthm, amssymb}
\usepackage{physics}
\usepackage{xcolor}
\usepackage{time}
\usepackage[pdftex,colorlinks=true]{hyperref}
\usepackage[colorinlistoftodos]{todonotes}
\usepackage{amsfonts}
\usepackage{mathtools}
\usepackage{bm}
\usepackage{comment}
\usepackage{ragged2e}
\usepackage{color,soul}
\usepackage{todonotes}
\newcounter{todocounter}

\def \be{\begin{equation}}
\def \ee{\end{equation}}
\def \bea{\begin{eqnarray}}
\def \eea{\end{eqnarray}}

\begin{document}
%
%
\title{Structure factor and topological bound of twisted bilayer semiconductors \\ at fractional fillings} 


\author{Timothy Zaklama}
\email{tzaklama@mit.edu}
\affiliation{Department of Physics, Massachusetts Institute of Technology, Cambridge, Massachusetts 02139, USA}

\author{Di Luo}
\affiliation{Department of Physics, Massachusetts Institute of Technology, Cambridge, Massachusetts 02139, USA}
\affiliation{The NSF AI Institute for Artificial Intelligence and Fundamental Interactions}
\affiliation{Department of Physics, Harvard University, Cambridge, MA 02138, USA}

\author{Liang Fu}
\email{liangfu@mit.edu}
\affiliation{Department of Physics, Massachusetts Institute of Technology, Cambridge, Massachusetts 02139, USA}


\begin{abstract}
The structure factor is a useful observable for probing charge density correlations in real materials, and its long-wavelength behavior encapsulated by ``quantum weight'' has recently gained prominence in the study of quantum geometry and topological phases of matter. Here we employ the full static structure factor, $S(\mathbf{q})$, to explore the phase diagram of twisted transition metal dichalcogenides (TMDs), specifically $t$MoTe$_2$, at filling factors $\nu=1/3$, $2/3$ under varying displacement fields. 
Our results reveal a topological phase transition between a fractional Chern insulator (FCI) and a generalized Wigner crystal (GWC). This transition is marked by the appearance of Bragg peaks at charge-density-wave vectors, and simultaneously, large decrease of $S(\mathbf{q})$ at small $\mathbf{q}$ which lowers the interaction energy. We further calculate the quantum weight of various FCI states, verifying the universal topological bound for interacting systems. Our findings provide new insights into the phase diagram of twisted TMDs and establish a general framework for characterizing topological phases through structure factor analysis.
\end{abstract}

\maketitle




{\it Introduction---.} The structure factor is a fundamental quantity for probing crystal structure and charge/spin density fluctuations in real materials, which can be measured by X-ray diffraction, electron loss spectroscopy and neutron scattering. In particular, the static (or equal-time) structure factor $S(\mathbf{q})$ is a ground state property defined as the Fourier transform of spatial density correlation. 
In periodic solids, Bragg peaks in the structure factor identify crystal structures and charge density wave orders \cite{girvin2019modern}.  
In quantum liquids, the static structure factor encodes useful information about the ground state and excitation spectrum \cite{nozieres1999theory}. For example, the structure factor of helium exhibits a characteristic peak at the wavevector set by the inverse particle distance, which is closely related to the roton excitation \cite{feynman1956energy}.

Recently, the behavior of the structure factor $S(\mathbf{q})$ at small $\mathbf{q}$, which characterizes long-wavelength density correlations, has  received growing attention. For gapped many-body systems, $S(\mathbf{q}\rightarrow 0)$ is generally quadratic in $\mathbf{q}$ and the quadratic coefficient $K$ defines a fundamental ground state property recently termed quantum weight \cite{onishi2023fundamental, onishi2024quantum}. Interestingly, quantum weight is directly related to optical response \cite{onishi2024quantum}, charge fluctuation \cite{estienne2022cornering, tam2024corner, wu2024corner} and many-body quantum geometry \cite{souza2000polarization}. It also sets an upper bound on the energy gap and a lower bound on the static dielectric constant of solids \cite{komissarov2022quantum, onishi2024quantum, souza2024optical}. 

Very recently, a universal lower bound for quantum weight has been established for Chern insulators \cite{onishi2024StructureFact}: $K \geq |C|$, where $C$ is the many-body Chern number. This inequality is derived from fundamental principles of physics and therefore 
applies generally to two-dimensional electron systems with either integral or fractional quantized Hall conductivity $\sigma_{xy} = C e^2/h$, with or without magnetic field. The bound on quantum weight is     
saturated in (integer and fractional) quantum Hall states that occur in a two-dimensional electron gas under strong magnetic fields \cite{girvin2019modern}, whereas the opposite behavior $K\gg |C|=1$ is found for the magnetic topological insulator MnBi$_2$Te$_4$ \cite{ghosh2024probing}.   

For noninteracting band insulators, the quantum weight $K$ is directly related to the spread of Wannier functions in real space \cite{marzari1997maximally}. Narrow-gap semiconductors have more extended Wannier functions and therefore larger $K$. 
For a Chern band, the bound $K\geq |C|$ dictates that the corresponding Wannier functions must have a minimum spread that is given by $C$ times lattice constant. More generally, the quantum weight of interacting two-dimensional systems is directly related to many-body quantum geometry, which is defined by the ground state wavefunction over twisted boundary conditions \cite{niu1985quantized, souza2000polarization}. While the geometry of Chern bands has been intensively studied in recent years \cite{roy2014band, claassen2015position, wang2021exact, mera2021relations, ledwith2022family,  Shavit2024quantum, qiu2024quantum, guerci2024layer}, little is known about the many-body quantum geometry of fractional Chern insulators (FCIs).   

In this work, we use band-projected exact diagonalization (ED) to calculate the full structure factor of twisted homobilayer transition metal dichalcogenides (TMDs) at various fractional fillings and displacement fields. Small-twist-angle bilayer TMDs host flat Chern bands \cite{wu2019topological, devakul2021magic}, which can enable robust ferromagnetism and fractional quantum anomalous Hall effect at fractional band fillings \cite{li2021spontaneous, crepel2023anomalous}. Recent experiments on twisted bilayer MoTe$_2$ have revealed a sequence of FCIs at zero magnetic field \cite{cai2023signatures, zeng2023thermodynamic, xu2023observation}, which bears a remarkable similarity with the Jain sequence fractional quantum Hall states in the lowest Landau level \cite{reddy2023fractional, wang2023fractional, goldman2023zero, dong2023composite, morales2023magic, crepel2023chiral, reddy2023toward, Yu2023fractional, abouelkomsan2024band, shi2024adiabatic, li2024variational}. Unlike the latter, however, theory predicts that $t$MoTe$_2$ also hosts generalized Wigner crystals (GWCs) \cite{reddy2023fractional, morales2023pressure, sheng2024topological,lu2024interaction}, anomalous Hall metals \cite{reddy2023toward}, and quantum anomalous Hall crystals \cite{sheng2024quantum, xu2023maximally}, leading to a fascinating phase diagram.   

Based on the structure factor, we identify and distinguish FCIs and GWCs at $\nu=1/3$ and $2/3$ fillings under small and large displacement fields respectively. The displacement field induced topological phase transition between the FCI and  GWC states at $\nu=2/3$ is marked by the abrupt decrease of quantum weight below the topological bound, which occurs in tandem with the emergence of charge-density-wave Bragg peaks in order to lower the interaction energy. 
Our work further reveals a magic angle where the topological bound for FCIs is nearly saturated and demonstrates the persistence of $\nu=2/3$ FCI to larger twist angles where the quantum weight significantly exceeds the bound.

\par

\begin{figure}[t]
    \includegraphics[width=0.48\textwidth]{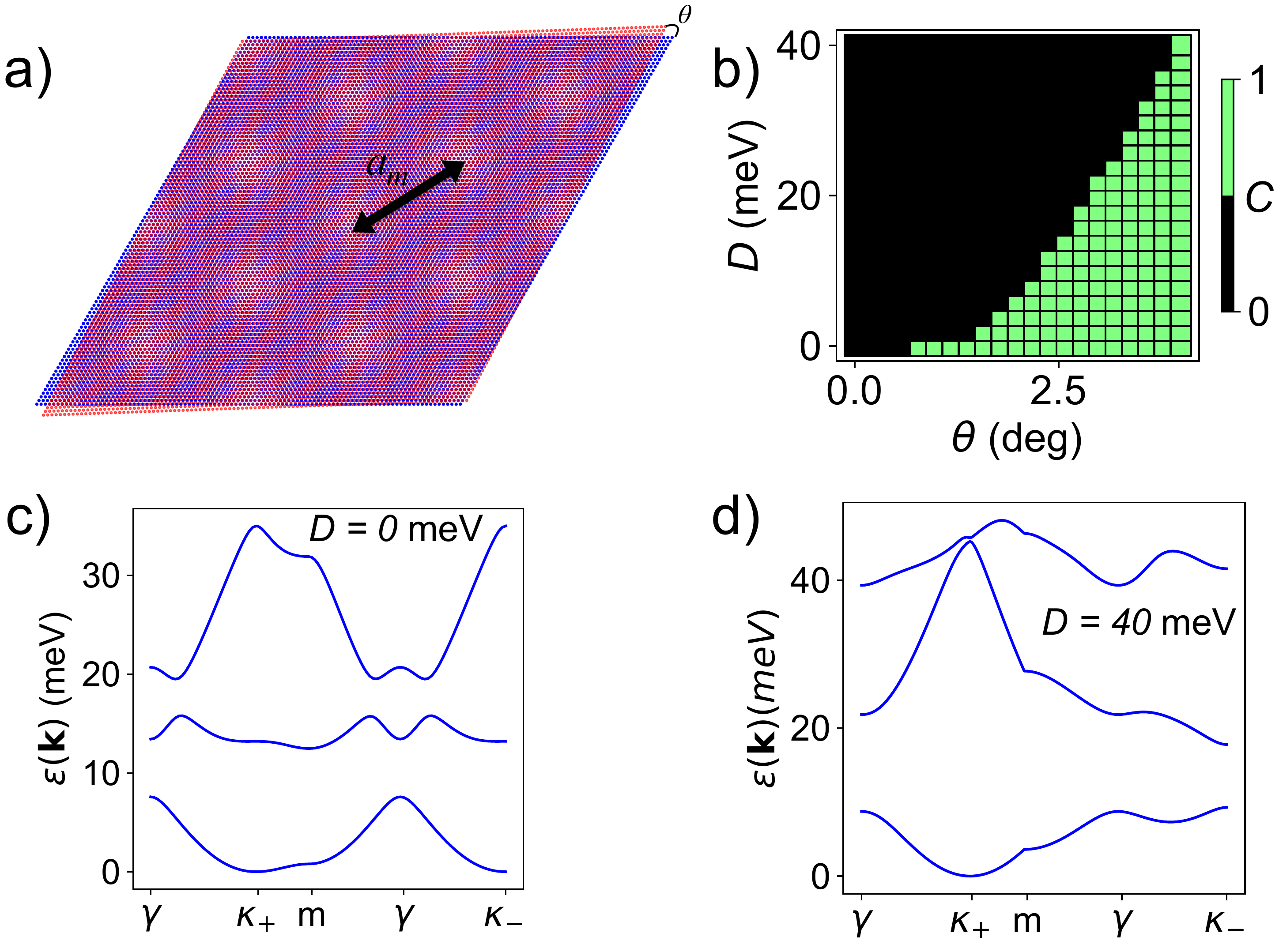} \label{fig:1}\quad \\\vskip 0.5cm
    \justifying
    \caption{\small Schematics and single particle picture. (a) Schematic of emergent moiré superlattice in real space and (b) the single particle Chern number as a function of displacement field and twist angle. There is a clear trend that as displacement field gets larger, one must go to larger twist angle to expect topology. (c, d) the continuum band structure for $\theta=2.8^\circ$ and $D=0,40$ meV, respectively.}
    \label{Fig1: Schematic and BS}
\end{figure}

{\it Moir\'e band structure and band topology---.} Our study is based on the interacting continuum model for holes in twisted TMD homobilayers schematically represented in Fig. \ref{Fig1: Schematic and BS} (a).  The single-particle Hamiltonian for spin-$1/2$ holes is given by \cite{wu2019topological}    
\begin{equation}
    \mathcal{H}_{\uparrow} = \begin{pmatrix}
  \frac{\hbar^2(-i\nabla -\kappa_+)^2}{2m^*} + \Tilde{V}_+(\mathbf{r})  & t(\mathbf{r}) \\ 
  t^\dagger(\mathbf{r}) & \frac{\hbar^2(-i\nabla-\kappa_-)^2}{2m^*}+ \Tilde{V}_-(\mathbf{r}) \end{pmatrix},
\end{equation}
while $\mathcal{H}_{\downarrow}$ is its time reversal conjugate. 
Here, $\mathbf{\kappa}_{\pm}=\frac{4\pi}{3a_M}\left(-\frac{\sqrt{3}}{2},\pm\frac{1}{2}\right)$ is introduced by rotational misalignment, 
with $a_M$ the moiré period and $m^*$ the effective mass. $V_{\pm}(\mathbf{r}) =  -2V\sum_{i=1,3,5} \cos(\mathbf{g}_i+\phi_\pm) \mp \frac{D}{2}$ denotes the moiré potential on each layer, where $D$ is the layer potential bias introduced by the displacement field, and $t(\mathbf{r})=w\left(1 + e^{-i\mathbf{g}_2\cdot\mathbf{r}} + e^{-i\mathbf{g}_3\cdot\mathbf{r}}\right)$ denotes the interlayer tunneling. The moiré reciprocal lattice vectors are $\mathbf{g}_i=\frac{4\pi}{\sqrt{3}a_M}\left(\cos(\frac{\pi(i-1)}{3}),\sin(\frac{\pi(i-1)}{3})\right)$ for $i=1,\dots,6$, with $\phi_+=-\phi_-=\phi$. 
We take the explicit forms of the parameters for $t$MoTe$_2$ from Ref. \cite{reddy2023fractional}.  

\par 

The displacement field tunes the dispersion, wavefunction and topology of moir\'e bands \cite{wu2019topological, devakul2021magic}. A large displacement field polarizes charges to one layer, and thereby drives the moir\'e bands from topological ($C=1$) to trivial ($C=0$), as shown in Fig. \ref{Fig1: Schematic and BS} (b).  
Fig. \ref{Fig1: Schematic and BS} (c) and (d) show the moir\'e band structure and the Chern number of the lowest band for two representative displacement fields $D=0$ and $40$ meV. 
This work will focus on the effect of displacement field on the ground state of twisted homobilayer TMDs at fractional fillings. 

{\it Band-Projected Exact Diagonalization---.}  The full continuum model Hamiltonian including electron-electron interaction is given by
\begin{equation}
\begin{split}
    \bm{H} &= \bm{H}_0 + \bm{V}, \\
    \bm{V} &= \frac{1}{2}\sum_{\sigma,\sigma'}\int d\mathbf{r}d\mathbf{r'}\psi_\sigma^\dagger(\mathbf{r})\psi_{\sigma'}^\dagger(\mathbf{r'})V(\mathbf{r}-\mathbf{r'})\psi_{\sigma'}(\mathbf{r'})\psi_{\sigma}(\mathbf{r}),
\end{split}
\end{equation}
where $\bm{H_0} = \sum_{\sigma=\uparrow,\downarrow}\int d\bm{r} \psi_\sigma^\dagger \mathcal{H}_\sigma \psi_\sigma$. Here we use a long-range Coulomb interaction $V(\mathbf{r})=\frac{e^2}{\epsilon r}$. By diagonalizing the one-body Hamiltonian $\bm{H}_0$, we obtain the band dispersion and Bloch wavefunction. 
Then, $\bm{H}$ can be rewritten in Bloch band basis as follows: 
\begin{multline}
    \bm{\Tilde{H}}=\sum_{\sigma,n,\mathbf{k}}\epsilon_{\sigma,n}(\mathbf{k})c_{\sigma,n,\mathbf{k}}^\dagger c_{\sigma,n,\mathbf{k}} \\
    +\frac{1}{2} \sum_{\substack{\sigma,\sigma'; \mathbf{k'_1}\\ n_1,n_2,n_3,n_4;\\ \mathbf{k'_2}\mathbf{k_1}, \mathbf{k_2}}} V_{\mathbf{k'_1}\mathbf{k'_2}\mathbf{k_1}\mathbf{k_2}}^{\sigma\sigma'} c^\dagger_{\sigma_1,n_1,\mathbf{k'_1}} c^\dagger_{\sigma_2,n_3,\mathbf{k'_2}} c_{\sigma_2,n_4,\mathbf{k_2}} c_{\sigma_1,n_2,\mathbf{k_1}} \label{Hband}
\end{multline}
where $c_{\sigma,n,\mathbf{k}}^\dagger$ creates a hole in a Bloch state at spin/valley $\sigma$, band $n$,  crystal momentum $\mathbf{k}$, which has a corresponding single-particle energy $\epsilon_{\sigma,n}(\mathbf{k})$. The Coulomb interaction matrix elements take on the Bloch basis representation $V_{\mathbf{k'_1}\mathbf{k'_2}\mathbf{k_1},\mathbf{k_2}}^{\sigma\sigma'} \equiv \langle \mathbf{k'_1}\sigma;\mathbf{k'_2}\sigma'|\hat{V}|\mathbf{k_1}\sigma;\mathbf{k_2}\sigma' \rangle$.

To solve many-body ground states at fractional fillings $\nu<1$ with exact diagonalization, we truncate the full Hilbert space to the subspace spanned by the lowest band, i.e., we only keep terms with $n_1=n_2=n_3=n_4=1$ in Eq. \eqref{Hband}. This band projection neglects band mixing with higher bands, which is accurate when the ratio of the characteristic Coulomb energy $\frac{e^2}{\epsilon a_M}$ to the moiré band gap is sufficiently small. While band mixing is quantitatively important for $t$MoTe$_2$, band-projected ED captures the essential physics of FCIs \cite{abouelkomsan2024band,Yu2023fractional}.

Further, we assume that the system at the range of fillings considered hereafter is fully spin-polarized, as found by previous numerical studies \cite{crepel2023anomalous, reddy2023fractional, Yu2023fractional}.
Our ED calculation uses the charge-$U(1)$ and spatial translational symmetries to diagonalize within the common eigenspace of $N_e$ and center of mass (CoM) crystal momentum. 
The ED at $\nu = 1/3$, $2/3$ is performed on a 27-unit cell cluster  with $C_6$ symmetry using periodic boundary conditions, as done in previous studies of this system \cite{reddy2023toward,abouelkomsan2024band}. 


{\it Structure Factor and Quantum Weight---.} The structure factor is defined as the Fourier transform of the static density-density correlation function:
\bea
    \chi (\mathbf{r}_i, \mathbf{r}_j) = \langle\rho(\mathbf{r}_i)\rho(\mathbf{r}_j)\rangle =\frac{1}{A}\sum_{\mathbf{q}}e^{i\mathbf{q}(\mathbf{r}_j-\mathbf{r}_i)}S(\mathbf{q}),
\eea
where $S(\mathbf{q})=\frac{1}{A}\langle\rho(\mathbf{q})\rho(\mathbf{-q})\rangle$ for general fermionic momentum-space density operators $\rho(\mathbf{q})=\sum_{\mathbf{k}}f^\dagger_{\mathbf{k}}f_{\mathbf{k}+\mathbf{q}}$ in the {\it plane wave} basis, $\langle \cdots \rangle$ denotes the expectation value over the many-body ground state, and system area $A=\frac{\sqrt{3}a_M^2 N}{2}$ for cluster size $N$. We take $S(\mathbf{q})$ over a single ground state rather than averaging over degenerate many-body ground states.

The structure factor defined above is the standard one, which applies to all electronic systems. In contrast, a different structure factor has been considered in the literature on quantum Hall states \cite{girvin1986magneto} and recently twisted TMDs \cite{morales2023pressure, reddy2023toward, dong2023composite, wolf2024intraband, mao2024low}, which is defined by the {\it projected density operator} in the Chern band instead of the bare density operator $\rho$. It is important to note that the projected structure factor differs from the standard one, even when band mixing is negligible. The difference between the two structure factors is evident from their contrasting behaviors at large $\mathbf{q}$. At $\mathbf{q}\rightarrow \infty$, the projected structure factor vanishes, while the standard one approaches 1. In this work, we study the standard structure factor of twisted TMDs, which has not been calculated before.     

As a ground state property, $S(\mathbf{q})$ can be computed with a variety of numerical methods including exact diagonalization, making it a powerful tool for probing strongly correlated and highly entangled systems. Recently, it was recognized that the long-wavelength behavior of the (standard) structure factor $S(\mathbf{q})$ is an important quantity of quantum many-body systems. For systems with an energy gap, $S(\mathbf{q}\rightarrow 0)$ takes the general form
\bea
    S(\mathbf{q})= \frac{K_{\alpha \beta}}{2\pi}q_\alpha q_\beta + \cdots,
\eea
where the summation over spatial indices $\alpha$ and $\beta$ is implied. In an interacting system, the quantum weight can also be formulated in terms of the quantum geometric tensor for the many-body ground state under twisted boundary conditions \cite{onishi2023fundamental}. Interestingly, the trace of $K$ has a {\it universal} lower bound determined by the many-body Chern number $C$ (or equivalently, the quantized Hall conductivity) \cite{onishi2024StructureFact}:


\begin{equation}
    K \equiv K_{xx} + K_{yy} \geq  |C|. 
\end{equation}
This bound is saturated in Landau levels of two-dimensional electron gas due to the Galilean invariance. For (FCIs) such as twisted TMDs at fractional fillings, it remains unknown how close the bound is to the actual quantum weight.

To calculate the structure factor in band-projected ED, we first transform the fermion operators $f_{\mathbf{k'}}$ indexed by spin $\sigma$, layer $X$, and {\it plane-wave} momentum $\mathbf{k'}$ to the fermion operators $c_{\sigma,n,\mathbf{k}}$  in band basis indexed by $\sigma$, band index $n$, {\it crystal momentum} $\mathbf{k}$, and reciprocal lattice vector $\mathbf{G}$ as

\bea
    f_{\sigma,X,\mathbf{k}+\mathbf{G}} = \sum_n u_{\sigma,n;\mathbf{G},X}(\mathbf{k})c_{\sigma,n,\mathbf{k}}, \label{ferm_ops}
\eea
where $u_{\sigma,n;\mathbf{G},X}$ are the Bloch wavefunctions obtained from solving the single-particle Hamiltonian.
Then, the full static structure factor can be expressed solely in terms of the bands in which the many body ground state resides, making numerical calculation far more tractable. The explicit calculation of the structure factor, along with its analytically simplified analogue, is provided in the Supplemental Material. \par

\begin{figure}[t]
    \includegraphics[width=.48\textwidth]{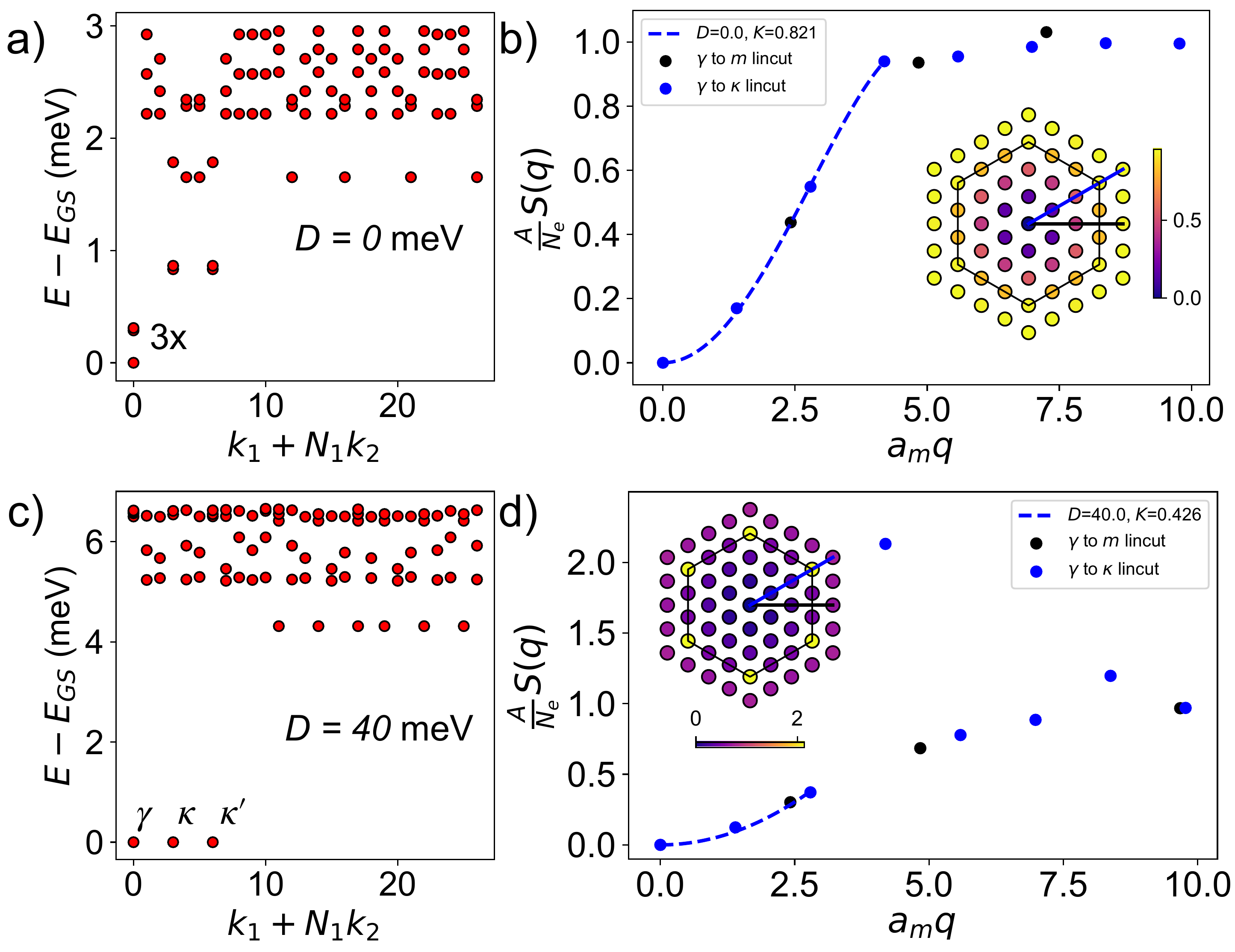} \quad \\\vskip 0.5cm
    \justifying
    \caption{\small FCI-to-GWC transition driven by displacement field. (a, c) Many Body Spectra on a 27 site cluster at $\nu=2/3$, $\theta=2.8^\circ$, $\epsilon=10$ and $D=0,40$ meV respectively. There is a clear near 3-fold degeneracy at $\gamma$ for $D=0$ meV that moves to $\gamma$, $\kappa$, $\kappa'$ for $D=40$ meV. (b, d) $\gamma$ to $\kappa$  and $\gamma$ to $M$ linecut of $S(\mathbf{q})$ with the corresponding parameters of (a, c). The 2d inset shows the full $S(\mathbf{q})$ one momentum shell outside of the Brillouin Zone and the points according to their corresponding linecuts.}
    \label{Fig2: MBS and S(q)}
\end{figure}

\par {\it Results---.} We now study the phase diagram of $t$MoTe$_2$ 
as displacement field is turned on. Focusing first on $\nu=2/3$ where the FCI state is most robust, we present in Fig. \ref{Fig2: MBS and S(q)} the many-body spectra (MBS) (the energy spectrum partitioned by momentum sector) and ground-state structure factor for $D=0$ and $D=40$ meV, where the lowest moir\'e band is separated from remote bands and has $C=1$ and $C=0$ respectively. 
Calculations were done at twist angle $\theta=2.8^\circ$  on a $N=27$ cluster with dielectric constant $\epsilon=10$.  

For $D=0$, we find three nearly-degenerate ground states in the same momentum sector, consistent with the previously identified  FCI phase on this cluster geometry \cite{reddy2023fractional}. 
In contrast, for $D=40$ meV, we find three degenerate ground states at distinct many-body momenta $\gamma, \kappa, \kappa'$, which is consistent with a $\sqrt{3}\times \sqrt{3}$ GWC \cite{sheng2024topological}. 

\begin{figure}[t]
    \includegraphics[width=.48\textwidth]{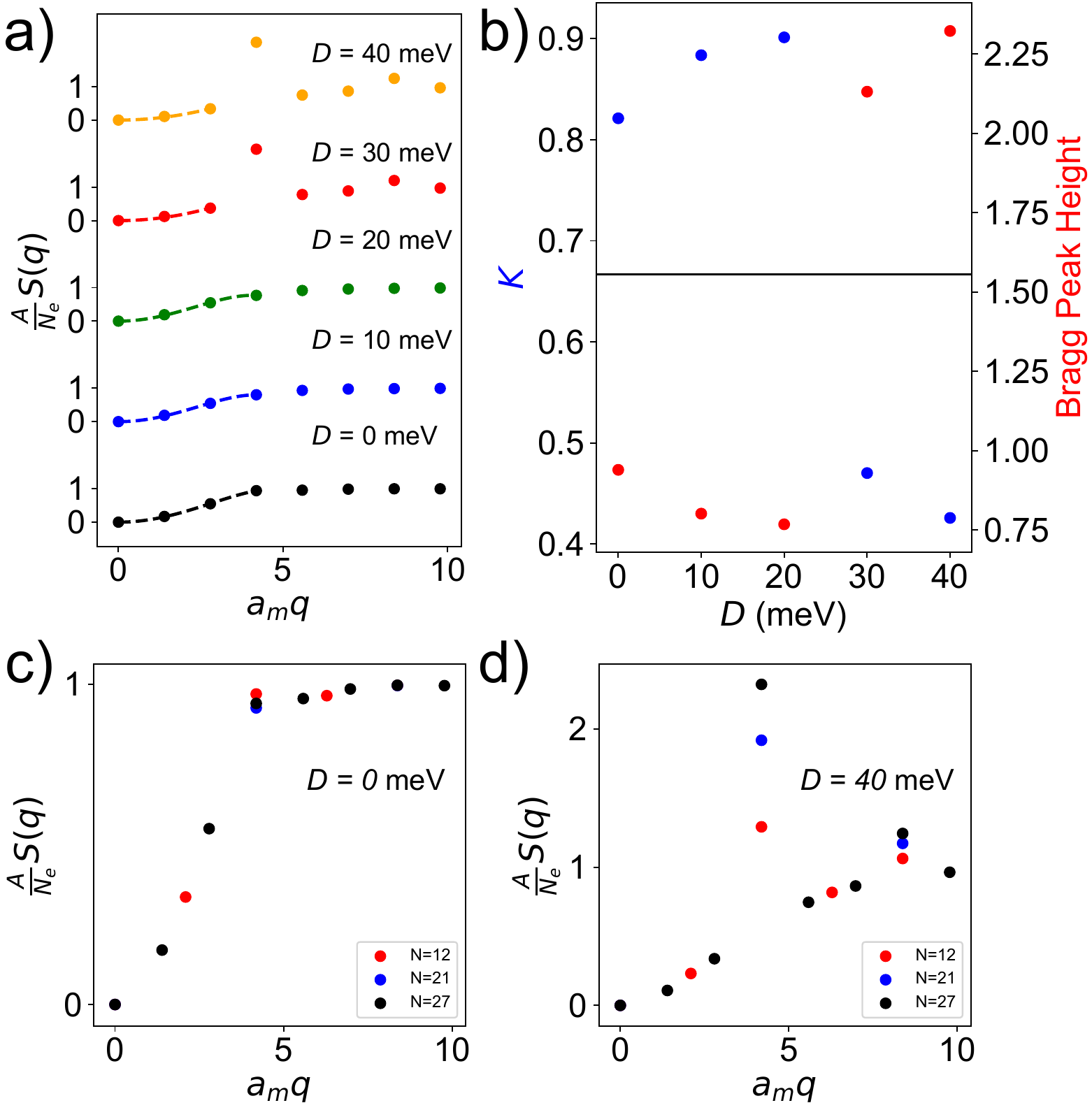} \quad \\\vskip 0.5cm
    \justifying
    \caption{\small $S(\mathbf{q})$ and quantum weight at various displacement fields for $\nu=2/3$, $\theta=2.8^\circ$ and $\epsilon=10$. (a) $S(\mathbf{q})$ of the $\gamma$ to $\kappa$ linecut $S(\mathbf{q})$ vertically displaced to show dependence on displacement field. Bragg peaks emerge at $\kappa$ for $D=30,40$ meV and scale accordingly. (b) Quantum Weight and  $S(\mathbf{q})$ value of $\kappa, \kappa'$ Bragg peak vs Displacement field. (c, d) $S(\mathbf{q})$ at increasing commensurate system size at $D=0,40$ meV respectively. Bragg peaks emerge and scale as system size in the GWC phase as expected.}
    \label{Fig3: S(q) and K}
\end{figure}

\begin{figure}[t]
    \includegraphics[width=.48\textwidth]{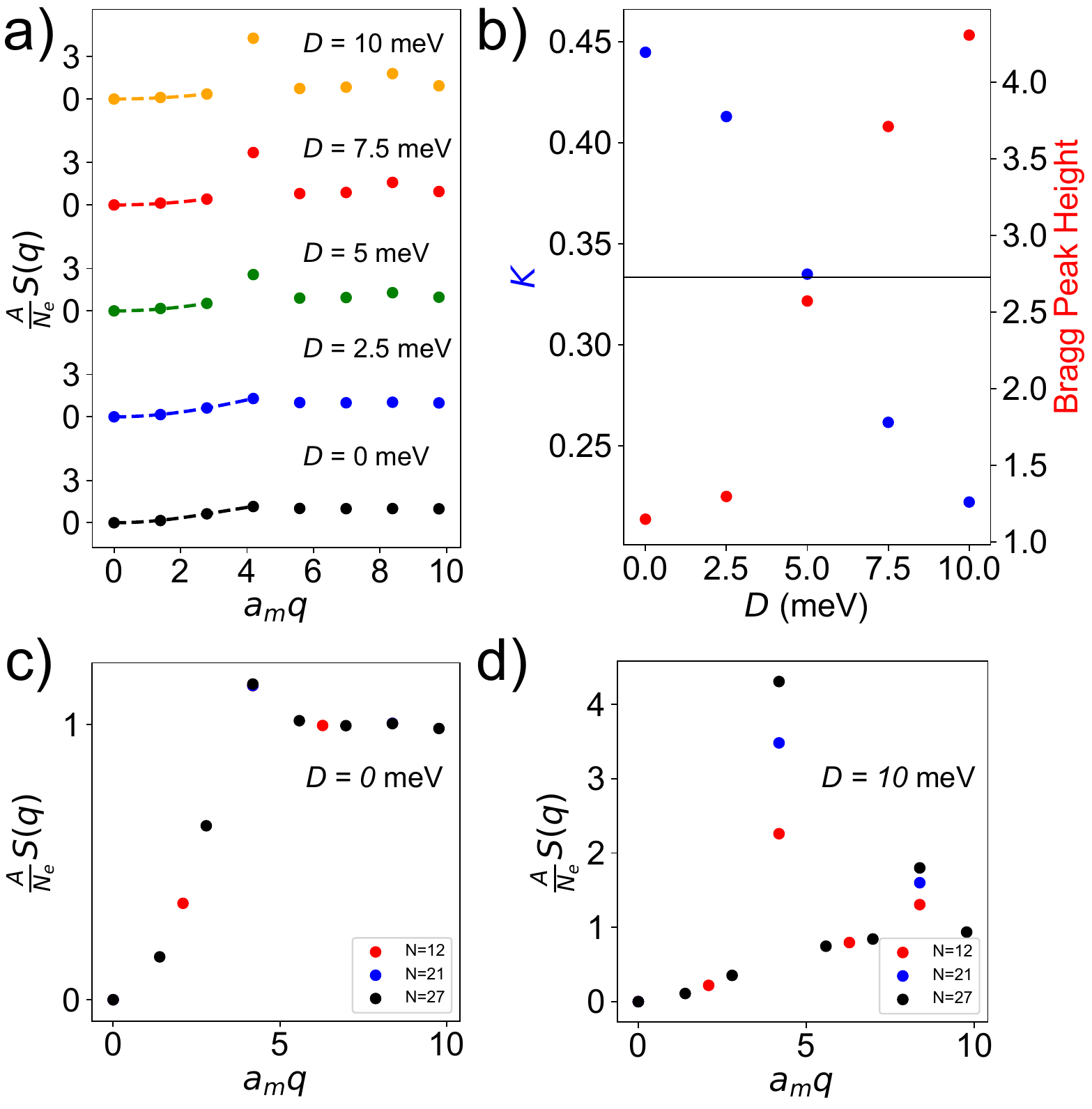} \quad \\\vskip 0.5cm
    \justifying
    \caption{\small $S(\mathbf{q})$ and quantum weight at various displacement fields for $\nu=1/3$, $\theta=2.0^\circ$ and $\epsilon=10$. (a) $S(\mathbf{q})$ of the $\gamma$ to $\kappa$ linecut similar to Fig. \ref{Fig2: MBS and S(q)}. (b) Quantum Weight and  $S(\mathbf{q})$ value of $\kappa, \kappa'$ Bragg peak vs Displacement field. (c, d) $S(\mathbf{q})$ at increasing commensurate system size at $D=0,10$ meV respectively. Bragg peaks emerge and scale as system size in the GWC phase as expected.}
    \label{Fig3_13: S(q) and K}
\end{figure}

The emergence of GWC that spontaneously breaks lattice translation symmetry can be diagnosed by the structure factor. Indeed, for $D=40$ meV, $S(\mathbf{q})$ shows a prominent peak at the expected wavevector.  
Moreover,  the peak height is found to increase with system size, which demonstrates the presence of long-range charge-density wave order, as shown in Fig. \ref{Fig3: S(q) and K} (c-d). 
The existence of $\sqrt{3}\times \sqrt{3}$  GWC is expected at large $D$, when charges reside on the MX moir\'e sites on one layer and form a honeycomb superstructure with tripled unit cell to minimize the Coulomb repulsion, similar to the case of TMD heterobilayers \cite{regan2020mott}. In contrast, the structure factor of the FCI state at $D=0$ is liquid like and qualitatively similar to that of the $\nu=2/3$ fractional quantum Hall state in the lowest Landau level.

We further analyze the structure factor $S(\mathbf{q})$ at small $\mathbf{q}$ and extract the quantum weight $K$ by a quadratic fitting shown in Fig. \ref{Fig2: MBS and S(q)} (b) and (d) (see Supplementary Material for details).  
We find $K=0.821$ at $D=0$, which indeed satisfies the universal topological bound of FCIs with $C=2/3$. 
In contrast, at $D=40$ meV, $K=0.426$ falls below $2/3$, which rules out the possibility of a $C=2/3$ FCI and instead is   
consistent with a topologically trivial state $C=0$.    
Our structure factor analysis shows that quantum weight can provide a useful method to 
distinguish topological and trivial states using a single ground state.  

We  extend this analysis further to identify the phase transition from FCI to GWC as the displacement field $D$ increases. Fig. \ref{Fig3: S(q) and K} (a) shows $S(\mathbf{q})$ along the direction $\gamma-\kappa$ at increasing $D$. From $D=0$ to 20 meV, $S(\mathbf{q})$ as a function of $\mathbf{q}$ increases smoothly from 0 to 1, characteristic of a quantum liquid.  At $D=30$ meV, a CDW Bragg peak arises abruptly and its magnitude increases further with $D$. Moreover, Fig. \ref{Fig3: S(q) and K} (b) shows that immediately when the Bragg peak develops, the $K$ falls below 2/3. Our findings provide strong evidence for a direct transition between FCI and GWC, which appears to be first-order as evidenced by the discontinuity in quantum weight.

Experiments on $t$MoTe$_2$ have indeed observed at $\nu=2/3$ a transition from FCI to a trivial insulating state ($\sigma_{xy}=0$) under increasing displacement field \cite{xu2023observation}. The latter is consistent with the GWC, whose $\sqrt{3}\times \sqrt{3}$ charge order may be detected by scanning tunneling microscopy, or identified through different exciton energy shifts at inequivalent moir\'e sites in the tripled unit cell of the GWC. 

We also study $t$MoTe$_2$ at $\nu=1/3$. Previous numerical studies at $D=0$ \cite{reddy2023fractional} have predicted that the $1/3$ state is a trivial $\sqrt{3} \times \sqrt{3}$ GWC at most of the twist angles, except in the neighborhood of a magic angle where the underlying Chern band is most lowest Landau level-like, giving rise to a $C=1/3$ FCI. In this case, our calculations find a similar FCI-to-GWC transition under increasing displacement field, as shown in Fig. \ref{Fig3_13: S(q) and K}. Compared to $\nu=2/3$, the FCI-to-GWC transition occurs at a smaller displacement field for $\nu=1/3$.

It is worth noting that across the $D$-induced transition from FCI to GWC at $\nu=2/3$ or $1/3$, the presence of Bragg peaks in the structure factor at CDW wavevector is accompanied by a reduction of $S(\mathbf{q})$ at small $\mathbf{q}$, i.e., a decrease of quantum weight. This is not a coincidence and can be heuristically understood from energetic considerations as follows: the transition to the GWC is expected to lower the interaction energy, which is directly related to the structure factor: $E_{\rm int} = \sum_{\mathbf{q}}V(\mathbf{q}) \left( \frac{A}{N_e}S(\mathbf{q})-1\right)/2A$, where Coulomb interaction $V(\mathbf{q}) = 2\pi e^2/\epsilon |\mathbf{q}|$ is positive at all $\mathbf{q}$, $N_e$ is the number of electrons in the system, and $A$ is the system area. On the other hand, the development of Bragg peaks in the GWC, however, indicates that $S(\mathbf{q})$ increases at the CDW wavevector. Therefore, the reduction of interaction energy can only be possible due to the decrease of $S(\mathbf{q})$ at other wavevectors, including small $\mathbf{q}$. We confirm the reduction of $E_{int}$ across the FCI-to-GWC transition in the Supplementary Material.

\begin{figure}[t]
    \includegraphics[width=0.48\textwidth]{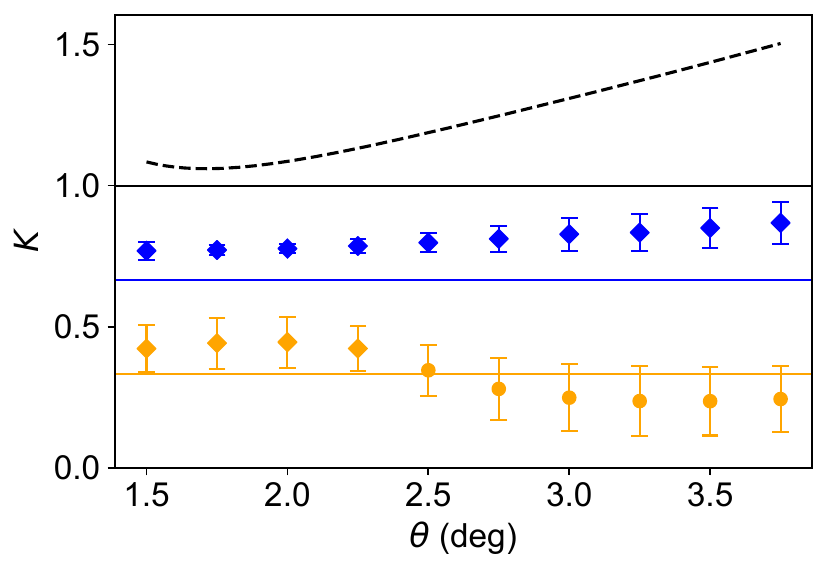} \label{fig:4}\quad \\\vskip 0.5cm
    \justifying
    \caption{\small Illustration of the universal topological bound being respected at fractional fillings. Extracted quantum weight is plotted versus twist angle at $\nu=1/3,2/3,1$ (orange, blue, black respectively). The universal topological bound for each filling is plotted as a solid horizontal line in the appropriate color, and the phase of the system at each angle is denoted by the style of the point. The diamonds correspond to points within the FCI phase, and the solid circles correspond to points within the GWC phase \cite{reddy2023fractional}. The dashed black line is the non-interacting case, previously shown to respect the topological bound in this system at integer filling \cite{onishi2024StructureFact}. All data is calculated at $\epsilon=5$ and $D=0$ on a 27 site cluster (error bars reflect the fit uncertainty due to finite size effects).} \label{Fig4: K vs Theta}
\end{figure}

Finally, we calculate the structure factor and extract the quantum weight at $D=0$ as a function of twist angle for fillings $\nu=1/3$ and $2/3$ in Fig. \ref{Fig4: K vs Theta}. 
For $\nu=1/3$, the system transitions from an FCI to a GWC at $\theta\approx 2.4^\circ$, as indicated by the emergence of Bragg peaks, in agreement with the previous study \cite{reddy2023fractional}. As in the case of the displacement-field induced FCI-GWC transition, we again see a fall in quantum weight below the topological bound across the phase boundary, signaling a topological phase transition. We find the mechanism for this shift is identical for the theta induced FCI-GWC transition: the emergence of Bragg peaks increases $S(\mathbf{q})$ at the CDW wavevector, driving $S(\mathbf{q})$ down for small $\mathbf{q}$, forcing the quantum weight below the topological lower bound. 

It is interesting to note that near the FCI-GWC phase boundary, the quantum weight nearly saturates the topological bound at a magic angle, which is around $2.4^\circ$ for $\nu=1/3$ and is less than $1.5^\circ$ for $\nu=2/3$. 
This is broadly consistent with the theoretical picture that the moir\'e band wavefunction at the magic angle closely resembles the lowest Landau level \cite{devakul2021magic, morales2023magic}, and therefore the  corresponding FCI states closely resembles the fractional quantum Hall states, whose quantum weight saturates the topological lower bound.   

Perhaps more remarkable is that the $\nu=2/3$ FCI persists to larger angles where the quantum weight far exceeds the topological bound. At $\theta=3.75^\circ$, $K=0.934$ is almost 3/2 times the topological bound 
signifying a considerable difference between 
this FCI state and the fractional quantum Hall state. It will be interesting to construct variational wavefunction for such non-standard FCI states featuring large quantum weight.         

\par {\it Discussion---.} In this work, we mapped out the phase diagram for $t$MoTe$_2$ at $\nu=1/3$ and $2/3$ as a function of displacement field based primarily on the structure factor calculated by band-projected ED. We use $S(\mathbf{q})$ to distinguish liquid and crystal states, and uncover a displacement field induced topological phase transition using the quantum weight $K$. 
Our numerical results confirm the topological bound $K \geq |C|$ found by Oinishi and Fu for FCIs and identify the magic angle where this bound is nearly saturated.   
\par 

We note that band-projected ED is quantitatively accurate when the band gap is large compared to interaction energy. However, near the critical field where the band gap closes and band topology changes $C=1 \rightarrow 0$, band mixing may affect the many-body ground states at fractional filling. Nonetheless, our band-projected ED provides variational ground state energies and wavefunctions, which serve as a useful starting point and benchmark for further investigation with advanced numerical methods. 

Our work has shown that the static structure factor encodes information about topological and correlated states in twisted TMDs. Further, our method is of broad and widespread application as it not only provides experimentalists with an additional tool to diagnose topology in real systems, but also theorists with a numerical means of measuring topology strictly from the ground state wavefunction. Novel numerical methods such as neural network variational Monte Carlo that do not calculate excited states have particular use for our method. In practice, measuring $S(\mathbf{q})$ directly with conventional x-ray scattering techniques is an experimental challenge due to the ultra-thin nature of 2D layers. In this regard, it is worth noting that the small-$q$ behavior of $S(\mathbf{q})$ encapsulated by the quantum weight can be directly determined from terahertz optical conductivity using the sum rule \cite{onishi2024quantum}: 
\begin{eqnarray}
K = 2\hbar \int_{0}^{\infty} d \omega \frac{ {\rm Re} \sigma_{xx}} {\omega},  
\end{eqnarray}
where the frequency integration should encompass the energy range of the continuum model for low-energy moir\'e bands.



\newpage

\par {\it Acknowledgement---.} We thank Yugo Onishi for related collaboration and Aidan Reddy and Ahmed Abouelkomsan for helpful discussions. We also acknowledge the github repository (https://github.com/AidanReddy/FermionED) that helped build the exact diagonalization and structure factor calculation code used in this work. This work was supported by the Air Force Office of Scientific Research under award number FA2386-24-1-4043. 
TZ was supported by the MIT Dean of Science Graduate Student Fellowship.
DL acknowledges support from the National Science Foundation under Cooperative Agreement PHY-2019786 (The NSF AI Institute for Artificial Intelligence and Fundamental Interactions,
http://iaifi.org/). LF was supported in part by a Simons Investigator Award from the Simons Foundation. 
\newpage\eject

\bibliography{ref}

\begin{thebibliography}{50}%
\makeatletter
\providecommand \@ifxundefined [1]{%
 \@ifx{#1\undefined}
}%
\providecommand \@ifnum [1]{%
 \ifnum #1\expandafter \@firstoftwo
 \else \expandafter \@secondoftwo
 \fi
}%
\providecommand \@ifx [1]{%
 \ifx #1\expandafter \@firstoftwo
 \else \expandafter \@secondoftwo
 \fi
}%
\providecommand \natexlab [1]{#1}%
\providecommand \enquote  [1]{``#1''}%
\providecommand \bibnamefont  [1]{#1}%
\providecommand \bibfnamefont [1]{#1}%
\providecommand \citenamefont [1]{#1}%
\providecommand \href@noop [0]{\@secondoftwo}%
\providecommand \href [0]{\begingroup \@sanitize@url \@href}%
\providecommand \@href[1]{\@@startlink{#1}\@@href}%
\providecommand \@@href[1]{\endgroup#1\@@endlink}%
\providecommand \@sanitize@url [0]{\catcode `\\12\catcode `\$12\catcode `\&12\catcode `\#12\catcode `\^12\catcode `\_12\catcode `\%12\relax}%
\providecommand \@@startlink[1]{}%
\providecommand \@@endlink[0]{}%
\providecommand \url  [0]{\begingroup\@sanitize@url \@url }%
\providecommand \@url [1]{\endgroup\@href {#1}{\urlprefix }}%
\providecommand \urlprefix  [0]{URL }%
\providecommand \Eprint [0]{\href }%
\providecommand \doibase [0]{https://doi.org/}%
\providecommand \selectlanguage [0]{\@gobble}%
\providecommand \bibinfo  [0]{\@secondoftwo}%
\providecommand \bibfield  [0]{\@secondoftwo}%
\providecommand \translation [1]{[#1]}%
\providecommand \BibitemOpen [0]{}%
\providecommand \bibitemStop [0]{}%
\providecommand \bibitemNoStop [0]{.\EOS\space}%
\providecommand \EOS [0]{\spacefactor3000\relax}%
\providecommand \BibitemShut  [1]{\csname bibitem#1\endcsname}%
\let\auto@bib@innerbib\@empty
\bibitem [{\citenamefont {Girvin}\ and\ \citenamefont {Yang}(2019)}]{girvin2019modern}%
  \BibitemOpen
  \bibfield  {author} {\bibinfo {author} {\bibfnamefont {S.~M.}\ \bibnamefont {Girvin}}\ and\ \bibinfo {author} {\bibfnamefont {K.}~\bibnamefont {Yang}},\ }\href@noop {} {\emph {\bibinfo {title} {Modern condensed matter physics}}}\ (\bibinfo  {publisher} {Cambridge University Press},\ \bibinfo {year} {2019})\BibitemShut {NoStop}%
\bibitem [{\citenamefont {Nozieres}\ and\ \citenamefont {Pines}(1999)}]{nozieres1999theory}%
  \BibitemOpen
  \bibfield  {author} {\bibinfo {author} {\bibfnamefont {P.}~\bibnamefont {Nozieres}}\ and\ \bibinfo {author} {\bibfnamefont {D.}~\bibnamefont {Pines}},\ }\href {https://books.google.com/books?id=q3wCwaV-gmUC} {\emph {\bibinfo {title} {Theory Of Quantum Liquids}}},\ Advanced Books Classics\ (\bibinfo  {publisher} {Avalon Publishing},\ \bibinfo {year} {1999})\BibitemShut {NoStop}%
\bibitem [{\citenamefont {Feynman}\ and\ \citenamefont {Cohen}(1956)}]{feynman1956energy}%
  \BibitemOpen
  \bibfield  {author} {\bibinfo {author} {\bibfnamefont {R.~P.}\ \bibnamefont {Feynman}}\ and\ \bibinfo {author} {\bibfnamefont {M.}~\bibnamefont {Cohen}},\ }\bibfield  {title} {\bibinfo {title} {Energy spectrum of the excitations in liquid helium},\ }\href {https://doi.org/10.1103/PhysRev.102.1189} {\bibfield  {journal} {\bibinfo  {journal} {Phys. Rev.}\ }\textbf {\bibinfo {volume} {102}},\ \bibinfo {pages} {1189} (\bibinfo {year} {1956})}\BibitemShut {NoStop}%
\bibitem [{\citenamefont {Onishi}\ and\ \citenamefont {Fu}(2024{\natexlab{a}})}]{onishi2023fundamental}%
  \BibitemOpen
  \bibfield  {author} {\bibinfo {author} {\bibfnamefont {Y.}~\bibnamefont {Onishi}}\ and\ \bibinfo {author} {\bibfnamefont {L.}~\bibnamefont {Fu}},\ }\bibfield  {title} {\bibinfo {title} {{Fundamental Bound on Topological Gap}},\ }\href {https://doi.org/10.1103/PhysRevX.14.011052} {\bibfield  {journal} {\bibinfo  {journal} {Phys. Rev. X}\ }\textbf {\bibinfo {volume} {14}},\ \bibinfo {pages} {011052} (\bibinfo {year} {2024}{\natexlab{a}})}\BibitemShut {NoStop}%
\bibitem [{\citenamefont {Onishi}\ and\ \citenamefont {Fu}(2024{\natexlab{b}})}]{onishi2024quantum}%
  \BibitemOpen
  \bibfield  {author} {\bibinfo {author} {\bibfnamefont {Y.}~\bibnamefont {Onishi}}\ and\ \bibinfo {author} {\bibfnamefont {L.}~\bibnamefont {Fu}},\ }\bibfield  {title} {\bibinfo {title} {Quantum weight},\ }\href@noop {} {\bibfield  {journal} {\bibinfo  {journal} {arXiv preprint arXiv:2406.06783}\ } (\bibinfo {year} {2024}{\natexlab{b}})}\BibitemShut {NoStop}%
\bibitem [{\citenamefont {Estienne}\ \emph {et~al.}(2022)\citenamefont {Estienne}, \citenamefont {Stéphan},\ and\ \citenamefont {Witczak-Krempa}}]{estienne2022cornering}%
  \BibitemOpen
  \bibfield  {author} {\bibinfo {author} {\bibfnamefont {B.}~\bibnamefont {Estienne}}, \bibinfo {author} {\bibfnamefont {J.}~\bibnamefont {Stéphan}},\ and\ \bibinfo {author} {\bibfnamefont {W.}~\bibnamefont {Witczak-Krempa}},\ }\bibfield  {title} {\bibinfo {title} {Cornering the universal shape of fluctuations},\ }\href {https://doi.org/10.1038/s41467-021-27727-1} {\bibfield  {journal} {\bibinfo  {journal} {Nature communications}\ }\textbf {\bibinfo {volume} {13}},\ \bibinfo {pages} {287} (\bibinfo {year} {2022})}\BibitemShut {NoStop}%
\bibitem [{\citenamefont {Tam}\ \emph {et~al.}(2024)\citenamefont {Tam}, \citenamefont {Herzog-Arbeitman},\ and\ \citenamefont {Yu}}]{tam2024corner}%
  \BibitemOpen
  \bibfield  {author} {\bibinfo {author} {\bibfnamefont {P.~M.}\ \bibnamefont {Tam}}, \bibinfo {author} {\bibfnamefont {J.}~\bibnamefont {Herzog-Arbeitman}},\ and\ \bibinfo {author} {\bibfnamefont {J.}~\bibnamefont {Yu}},\ }\href {https://arxiv.org/abs/2406.17023} {\bibinfo {title} {Corner charge fluctuation as an observable for quantum geometry and entanglement in two-dimensional insulators}} (\bibinfo {year} {2024}),\ \Eprint {https://arxiv.org/abs/2406.17023} {arXiv:2406.17023 [cond-mat.mes-hall]} \BibitemShut {NoStop}%
\bibitem [{\citenamefont {Wu}\ \emph {et~al.}(2024)\citenamefont {Wu}, \citenamefont {Cai}, \citenamefont {Cheng},\ and\ \citenamefont {Kumar}}]{wu2024corner}%
  \BibitemOpen
  \bibfield  {author} {\bibinfo {author} {\bibfnamefont {X.-C.}\ \bibnamefont {Wu}}, \bibinfo {author} {\bibfnamefont {K.-L.}\ \bibnamefont {Cai}}, \bibinfo {author} {\bibfnamefont {M.}~\bibnamefont {Cheng}},\ and\ \bibinfo {author} {\bibfnamefont {P.}~\bibnamefont {Kumar}},\ }\href {https://arxiv.org/abs/2408.16057} {\bibinfo {title} {Corner charge fluctuations and many-body quantum geometry}} (\bibinfo {year} {2024}),\ \Eprint {https://arxiv.org/abs/2408.16057} {arXiv:2408.16057 [cond-mat.str-el]} \BibitemShut {NoStop}%
\bibitem [{\citenamefont {Souza}\ \emph {et~al.}(2000)\citenamefont {Souza}, \citenamefont {Wilkens},\ and\ \citenamefont {Martin}}]{souza2000polarization}%
  \BibitemOpen
  \bibfield  {author} {\bibinfo {author} {\bibfnamefont {I.}~\bibnamefont {Souza}}, \bibinfo {author} {\bibfnamefont {T.}~\bibnamefont {Wilkens}},\ and\ \bibinfo {author} {\bibfnamefont {R.~M.}\ \bibnamefont {Martin}},\ }\bibfield  {title} {\bibinfo {title} {Polarization and localization in insulators: Generating function approach},\ }\href {https://doi.org/10.1103/PhysRevB.62.1666} {\bibfield  {journal} {\bibinfo  {journal} {Phys. Rev. B}\ }\textbf {\bibinfo {volume} {62}},\ \bibinfo {pages} {1666} (\bibinfo {year} {2000})}\BibitemShut {NoStop}%
\bibitem [{\citenamefont {Komissarov}\ \emph {et~al.}(2024)\citenamefont {Komissarov}, \citenamefont {Holder},\ and\ \citenamefont {Queiroz}}]{komissarov2022quantum}%
  \BibitemOpen
  \bibfield  {author} {\bibinfo {author} {\bibfnamefont {I.}~\bibnamefont {Komissarov}}, \bibinfo {author} {\bibfnamefont {T.}~\bibnamefont {Holder}},\ and\ \bibinfo {author} {\bibfnamefont {R.}~\bibnamefont {Queiroz}},\ }\bibfield  {title} {\bibinfo {title} {The quantum geometric origin of capacitance in insulators},\ }\href {https://doi.org/10.1038/s41467-024-48808-x} {\bibfield  {journal} {\bibinfo  {journal} {Nature communications}\ }\textbf {\bibinfo {volume} {15}},\ \bibinfo {pages} {4621} (\bibinfo {year} {2024})}\BibitemShut {NoStop}%
\bibitem [{\citenamefont {Souza}\ \emph {et~al.}(2024)\citenamefont {Souza}, \citenamefont {Martin},\ and\ \citenamefont {Stengel}}]{souza2024optical}%
  \BibitemOpen
  \bibfield  {author} {\bibinfo {author} {\bibfnamefont {I.}~\bibnamefont {Souza}}, \bibinfo {author} {\bibfnamefont {R.~M.}\ \bibnamefont {Martin}},\ and\ \bibinfo {author} {\bibfnamefont {M.}~\bibnamefont {Stengel}},\ }\href {https://arxiv.org/abs/2407.17908} {\bibinfo {title} {Optical bounds on many-electron localization}} (\bibinfo {year} {2024}),\ \Eprint {https://arxiv.org/abs/2407.17908} {arXiv:2407.17908 [cond-mat.mtrl-sci]} \BibitemShut {NoStop}%
\bibitem [{\citenamefont {Onishi}\ and\ \citenamefont {Fu}(2024{\natexlab{c}})}]{onishi2024StructureFact}%
  \BibitemOpen
  \bibfield  {author} {\bibinfo {author} {\bibfnamefont {Y.}~\bibnamefont {Onishi}}\ and\ \bibinfo {author} {\bibfnamefont {L.}~\bibnamefont {Fu}},\ }\bibfield  {title} {\bibinfo {title} {Topological bound on the structure factor},\ }\href {https://doi.org/10.1103/PhysRevLett.133.206602} {\bibfield  {journal} {\bibinfo  {journal} {Phys. Rev. Lett.}\ }\textbf {\bibinfo {volume} {133}},\ \bibinfo {pages} {206602} (\bibinfo {year} {2024}{\natexlab{c}})}\BibitemShut {NoStop}%
\bibitem [{\citenamefont {Ghosh}\ \emph {et~al.}(2024)\citenamefont {Ghosh}, \citenamefont {Onishi}, \citenamefont {Xu}, \citenamefont {Lin}, \citenamefont {Fu},\ and\ \citenamefont {Bansil}}]{ghosh2024probing}%
  \BibitemOpen
  \bibfield  {author} {\bibinfo {author} {\bibfnamefont {B.}~\bibnamefont {Ghosh}}, \bibinfo {author} {\bibfnamefont {Y.}~\bibnamefont {Onishi}}, \bibinfo {author} {\bibfnamefont {S.-Y.}\ \bibnamefont {Xu}}, \bibinfo {author} {\bibfnamefont {H.}~\bibnamefont {Lin}}, \bibinfo {author} {\bibfnamefont {L.}~\bibnamefont {Fu}},\ and\ \bibinfo {author} {\bibfnamefont {A.}~\bibnamefont {Bansil}},\ }\href {https://arxiv.org/abs/2401.09689} {\bibinfo {title} {Probing quantum geometry through optical conductivity and magnetic circular dichroism}} (\bibinfo {year} {2024}),\ \Eprint {https://arxiv.org/abs/2401.09689} {arXiv:2401.09689 [cond-mat.mtrl-sci]} \BibitemShut {NoStop}%
\bibitem [{\citenamefont {Marzari}\ and\ \citenamefont {Vanderbilt}(1997)}]{marzari1997maximally}%
  \BibitemOpen
  \bibfield  {author} {\bibinfo {author} {\bibfnamefont {N.}~\bibnamefont {Marzari}}\ and\ \bibinfo {author} {\bibfnamefont {D.}~\bibnamefont {Vanderbilt}},\ }\bibfield  {title} {\bibinfo {title} {Maximally localized generalized wannier functions for composite energy bands},\ }\href {https://doi.org/10.1103/PhysRevB.56.12847} {\bibfield  {journal} {\bibinfo  {journal} {Phys. Rev. B}\ }\textbf {\bibinfo {volume} {56}},\ \bibinfo {pages} {12847} (\bibinfo {year} {1997})}\BibitemShut {NoStop}%
\bibitem [{\citenamefont {Niu}\ \emph {et~al.}(1985)\citenamefont {Niu}, \citenamefont {Thouless},\ and\ \citenamefont {Wu}}]{niu1985quantized}%
  \BibitemOpen
  \bibfield  {author} {\bibinfo {author} {\bibfnamefont {Q.}~\bibnamefont {Niu}}, \bibinfo {author} {\bibfnamefont {D.~J.}\ \bibnamefont {Thouless}},\ and\ \bibinfo {author} {\bibfnamefont {Y.-S.}\ \bibnamefont {Wu}},\ }\bibfield  {title} {\bibinfo {title} {Quantized hall conductance as a topological invariant},\ }\href {https://doi.org/10.1103/PhysRevB.31.3372} {\bibfield  {journal} {\bibinfo  {journal} {Phys. Rev. B}\ }\textbf {\bibinfo {volume} {31}},\ \bibinfo {pages} {3372} (\bibinfo {year} {1985})}\BibitemShut {NoStop}%
\bibitem [{\citenamefont {Roy}(2014)}]{roy2014band}%
  \BibitemOpen
  \bibfield  {author} {\bibinfo {author} {\bibfnamefont {R.}~\bibnamefont {Roy}},\ }\bibfield  {title} {\bibinfo {title} {Band geometry of fractional topological insulators},\ }\href@noop {} {\bibfield  {journal} {\bibinfo  {journal} {Physical Review B}\ }\textbf {\bibinfo {volume} {90}},\ \bibinfo {pages} {165139} (\bibinfo {year} {2014})}\BibitemShut {NoStop}%
\bibitem [{\citenamefont {Claassen}\ \emph {et~al.}(2015)\citenamefont {Claassen}, \citenamefont {Lee}, \citenamefont {Thomale}, \citenamefont {Qi},\ and\ \citenamefont {Devereaux}}]{claassen2015position}%
  \BibitemOpen
  \bibfield  {author} {\bibinfo {author} {\bibfnamefont {M.}~\bibnamefont {Claassen}}, \bibinfo {author} {\bibfnamefont {C.~H.}\ \bibnamefont {Lee}}, \bibinfo {author} {\bibfnamefont {R.}~\bibnamefont {Thomale}}, \bibinfo {author} {\bibfnamefont {X.-L.}\ \bibnamefont {Qi}},\ and\ \bibinfo {author} {\bibfnamefont {T.~P.}\ \bibnamefont {Devereaux}},\ }\bibfield  {title} {\bibinfo {title} {Position-momentum duality and fractional quantum hall effect in chern insulators},\ }\href {https://doi.org/10.1103/PhysRevLett.114.236802} {\bibfield  {journal} {\bibinfo  {journal} {Phys. Rev. Lett.}\ }\textbf {\bibinfo {volume} {114}},\ \bibinfo {pages} {236802} (\bibinfo {year} {2015})}\BibitemShut {NoStop}%
\bibitem [{\citenamefont {Wang}\ \emph {et~al.}(2021)\citenamefont {Wang}, \citenamefont {Cano}, \citenamefont {Millis}, \citenamefont {Liu},\ and\ \citenamefont {Yang}}]{wang2021exact}%
  \BibitemOpen
  \bibfield  {author} {\bibinfo {author} {\bibfnamefont {J.}~\bibnamefont {Wang}}, \bibinfo {author} {\bibfnamefont {J.}~\bibnamefont {Cano}}, \bibinfo {author} {\bibfnamefont {A.~J.}\ \bibnamefont {Millis}}, \bibinfo {author} {\bibfnamefont {Z.}~\bibnamefont {Liu}},\ and\ \bibinfo {author} {\bibfnamefont {B.}~\bibnamefont {Yang}},\ }\bibfield  {title} {\bibinfo {title} {Exact landau level description of geometry and interaction in a flatband},\ }\href@noop {} {\bibfield  {journal} {\bibinfo  {journal} {Physical review letters}\ }\textbf {\bibinfo {volume} {127}},\ \bibinfo {pages} {246403} (\bibinfo {year} {2021})}\BibitemShut {NoStop}%
\bibitem [{\citenamefont {Ozawa}\ and\ \citenamefont {Mera}(2021)}]{mera2021relations}%
  \BibitemOpen
  \bibfield  {author} {\bibinfo {author} {\bibfnamefont {T.}~\bibnamefont {Ozawa}}\ and\ \bibinfo {author} {\bibfnamefont {B.}~\bibnamefont {Mera}},\ }\bibfield  {title} {\bibinfo {title} {Relations between topology and the quantum metric for chern insulators},\ }\href {https://doi.org/10.1103/PhysRevB.104.045103} {\bibfield  {journal} {\bibinfo  {journal} {Phys. Rev. B}\ }\textbf {\bibinfo {volume} {104}},\ \bibinfo {pages} {045103} (\bibinfo {year} {2021})}\BibitemShut {NoStop}%
\bibitem [{\citenamefont {Ledwith}\ \emph {et~al.}(2022)\citenamefont {Ledwith}, \citenamefont {Vishwanath},\ and\ \citenamefont {Khalaf}}]{ledwith2022family}%
  \BibitemOpen
  \bibfield  {author} {\bibinfo {author} {\bibfnamefont {P.~J.}\ \bibnamefont {Ledwith}}, \bibinfo {author} {\bibfnamefont {A.}~\bibnamefont {Vishwanath}},\ and\ \bibinfo {author} {\bibfnamefont {E.}~\bibnamefont {Khalaf}},\ }\bibfield  {title} {\bibinfo {title} {Family of ideal chern flatbands with arbitrary chern number in chiral twisted graphene multilayers},\ }\href {https://doi.org/10.1103/PhysRevLett.128.176404} {\bibfield  {journal} {\bibinfo  {journal} {Phys. Rev. Lett.}\ }\textbf {\bibinfo {volume} {128}},\ \bibinfo {pages} {176404} (\bibinfo {year} {2022})}\BibitemShut {NoStop}%
\bibitem [{\citenamefont {Shavit}\ and\ \citenamefont {Oreg}(2024)}]{Shavit2024quantum}%
  \BibitemOpen
  \bibfield  {author} {\bibinfo {author} {\bibfnamefont {G.}~\bibnamefont {Shavit}}\ and\ \bibinfo {author} {\bibfnamefont {Y.}~\bibnamefont {Oreg}},\ }\bibfield  {title} {\bibinfo {title} {Quantum geometry and stabilization of fractional chern insulators far from the ideal limit},\ }\href {https://doi.org/10.1103/PhysRevLett.133.156504} {\bibfield  {journal} {\bibinfo  {journal} {Phys. Rev. Lett.}\ }\textbf {\bibinfo {volume} {133}},\ \bibinfo {pages} {156504} (\bibinfo {year} {2024})}\BibitemShut {NoStop}%
\bibitem [{\citenamefont {Qiu}\ and\ \citenamefont {Wu}(2024)}]{qiu2024quantum}%
  \BibitemOpen
  \bibfield  {author} {\bibinfo {author} {\bibfnamefont {W.-X.}\ \bibnamefont {Qiu}}\ and\ \bibinfo {author} {\bibfnamefont {F.}~\bibnamefont {Wu}},\ }\href {https://arxiv.org/abs/2407.03317} {\bibinfo {title} {Quantum geometry probed by chiral excitonic optical response of chern insulators}} (\bibinfo {year} {2024}),\ \Eprint {https://arxiv.org/abs/2407.03317} {arXiv:2407.03317 [cond-mat.mes-hall]} \BibitemShut {NoStop}%
\bibitem [{\citenamefont {Guerci}\ \emph {et~al.}(2024)\citenamefont {Guerci}, \citenamefont {Wang},\ and\ \citenamefont {Mora}}]{guerci2024layer}%
  \BibitemOpen
  \bibfield  {author} {\bibinfo {author} {\bibfnamefont {D.}~\bibnamefont {Guerci}}, \bibinfo {author} {\bibfnamefont {J.}~\bibnamefont {Wang}},\ and\ \bibinfo {author} {\bibfnamefont {C.}~\bibnamefont {Mora}},\ }\href {https://arxiv.org/abs/2408.12652} {\bibinfo {title} {Layer skyrmions for ideal chern bands and twisted bilayer graphene}} (\bibinfo {year} {2024}),\ \Eprint {https://arxiv.org/abs/2408.12652} {arXiv:2408.12652 [cond-mat.mes-hall]} \BibitemShut {NoStop}%
\bibitem [{\citenamefont {Wu}\ \emph {et~al.}(2019)\citenamefont {Wu}, \citenamefont {Lovorn}, \citenamefont {Tutuc}, \citenamefont {Martin},\ and\ \citenamefont {MacDonald}}]{wu2019topological}%
  \BibitemOpen
  \bibfield  {author} {\bibinfo {author} {\bibfnamefont {F.}~\bibnamefont {Wu}}, \bibinfo {author} {\bibfnamefont {T.}~\bibnamefont {Lovorn}}, \bibinfo {author} {\bibfnamefont {E.}~\bibnamefont {Tutuc}}, \bibinfo {author} {\bibfnamefont {I.}~\bibnamefont {Martin}},\ and\ \bibinfo {author} {\bibfnamefont {A.}~\bibnamefont {MacDonald}},\ }\bibfield  {title} {\bibinfo {title} {Topological insulators in twisted transition metal dichalcogenide homobilayers},\ }\href {https://doi.org/10.1103/PhysRevLett.122.086402} {\bibfield  {journal} {\bibinfo  {journal} {Physical review letters}\ }\textbf {\bibinfo {volume} {122}},\ \bibinfo {pages} {086402} (\bibinfo {year} {2019})}\BibitemShut {NoStop}%
\bibitem [{\citenamefont {Devakul}\ \emph {et~al.}(2021)\citenamefont {Devakul}, \citenamefont {Cr{\'e}pel}, \citenamefont {Zhang},\ and\ \citenamefont {Fu}}]{devakul2021magic}%
  \BibitemOpen
  \bibfield  {author} {\bibinfo {author} {\bibfnamefont {T.}~\bibnamefont {Devakul}}, \bibinfo {author} {\bibfnamefont {V.}~\bibnamefont {Cr{\'e}pel}}, \bibinfo {author} {\bibfnamefont {Y.}~\bibnamefont {Zhang}},\ and\ \bibinfo {author} {\bibfnamefont {L.}~\bibnamefont {Fu}},\ }\bibfield  {title} {\bibinfo {title} {Magic in twisted transition metal dichalcogenide bilayers},\ }\href {https://doi.org/10.1038/s41467-021-27042-9} {\bibfield  {journal} {\bibinfo  {journal} {Nature communications}\ }\textbf {\bibinfo {volume} {12}},\ \bibinfo {pages} {6730} (\bibinfo {year} {2021})}\BibitemShut {NoStop}%
\bibitem [{\citenamefont {Li}\ \emph {et~al.}(2021)\citenamefont {Li}, \citenamefont {Kumar}, \citenamefont {Sun},\ and\ \citenamefont {Lin}}]{li2021spontaneous}%
  \BibitemOpen
  \bibfield  {author} {\bibinfo {author} {\bibfnamefont {H.}~\bibnamefont {Li}}, \bibinfo {author} {\bibfnamefont {U.}~\bibnamefont {Kumar}}, \bibinfo {author} {\bibfnamefont {K.}~\bibnamefont {Sun}},\ and\ \bibinfo {author} {\bibfnamefont {S.-Z.}\ \bibnamefont {Lin}},\ }\bibfield  {title} {\bibinfo {title} {Spontaneous fractional chern insulators in transition metal dichalcogenide moir{\'e} superlattices},\ }\href {https://doi.org/10.1103/PhysRevResearch.3.L032070} {\bibfield  {journal} {\bibinfo  {journal} {Physical Review Research}\ }\textbf {\bibinfo {volume} {3}},\ \bibinfo {pages} {L032070} (\bibinfo {year} {2021})}\BibitemShut {NoStop}%
\bibitem [{\citenamefont {Cr{\'e}pel}\ and\ \citenamefont {Fu}(2023)}]{crepel2023anomalous}%
  \BibitemOpen
  \bibfield  {author} {\bibinfo {author} {\bibfnamefont {V.}~\bibnamefont {Cr{\'e}pel}}\ and\ \bibinfo {author} {\bibfnamefont {L.}~\bibnamefont {Fu}},\ }\bibfield  {title} {\bibinfo {title} {Anomalous hall metal and fractional chern insulator in twisted transition metal dichalcogenides},\ }\href@noop {} {\bibfield  {journal} {\bibinfo  {journal} {Physical Review B}\ }\textbf {\bibinfo {volume} {107}},\ \bibinfo {pages} {L201109} (\bibinfo {year} {2023})}\BibitemShut {NoStop}%
\bibitem [{\citenamefont {Cai}\ \emph {et~al.}(2023)\citenamefont {Cai}, \citenamefont {Anderson}, \citenamefont {Wang}, \citenamefont {Zhang}, \citenamefont {Liu}, \citenamefont {Holtzmann}, \citenamefont {Zhang}, \citenamefont {Fan}, \citenamefont {Taniguchi}, \citenamefont {Watanabe} \emph {et~al.}}]{cai2023signatures}%
  \BibitemOpen
  \bibfield  {author} {\bibinfo {author} {\bibfnamefont {J.}~\bibnamefont {Cai}}, \bibinfo {author} {\bibfnamefont {E.}~\bibnamefont {Anderson}}, \bibinfo {author} {\bibfnamefont {C.}~\bibnamefont {Wang}}, \bibinfo {author} {\bibfnamefont {X.}~\bibnamefont {Zhang}}, \bibinfo {author} {\bibfnamefont {X.}~\bibnamefont {Liu}}, \bibinfo {author} {\bibfnamefont {W.}~\bibnamefont {Holtzmann}}, \bibinfo {author} {\bibfnamefont {Y.}~\bibnamefont {Zhang}}, \bibinfo {author} {\bibfnamefont {F.}~\bibnamefont {Fan}}, \bibinfo {author} {\bibfnamefont {T.}~\bibnamefont {Taniguchi}}, \bibinfo {author} {\bibfnamefont {K.}~\bibnamefont {Watanabe}}, \emph {et~al.},\ }\bibfield  {title} {\bibinfo {title} {Signatures of fractional quantum anomalous hall states in twisted mote2},\ }\href {https://doi.org/10.1038/s41586-023-06289-w} {\bibfield  {journal} {\bibinfo  {journal} {Nature}\ ,\ \bibinfo {pages} {1}} (\bibinfo {year} {2023})}\BibitemShut {NoStop}%
\bibitem [{\citenamefont {Zeng}\ \emph {et~al.}(2023)\citenamefont {Zeng}, \citenamefont {Xia}, \citenamefont {Kang}, \citenamefont {Zhu}, \citenamefont {Kn{\"u}ppel}, \citenamefont {Vaswani}, \citenamefont {Watanabe}, \citenamefont {Taniguchi}, \citenamefont {Mak},\ and\ \citenamefont {Shan}}]{zeng2023thermodynamic}%
  \BibitemOpen
  \bibfield  {author} {\bibinfo {author} {\bibfnamefont {Y.}~\bibnamefont {Zeng}}, \bibinfo {author} {\bibfnamefont {Z.}~\bibnamefont {Xia}}, \bibinfo {author} {\bibfnamefont {K.}~\bibnamefont {Kang}}, \bibinfo {author} {\bibfnamefont {J.}~\bibnamefont {Zhu}}, \bibinfo {author} {\bibfnamefont {P.}~\bibnamefont {Kn{\"u}ppel}}, \bibinfo {author} {\bibfnamefont {C.}~\bibnamefont {Vaswani}}, \bibinfo {author} {\bibfnamefont {K.}~\bibnamefont {Watanabe}}, \bibinfo {author} {\bibfnamefont {T.}~\bibnamefont {Taniguchi}}, \bibinfo {author} {\bibfnamefont {K.~F.}\ \bibnamefont {Mak}},\ and\ \bibinfo {author} {\bibfnamefont {J.}~\bibnamefont {Shan}},\ }\bibfield  {title} {\bibinfo {title} {Thermodynamic evidence of fractional chern insulator in moir{\'e} mote2},\ }\href {https://www.nature.com/articles/s41586-023-06452-3} {\bibfield  {journal} {\bibinfo  {journal} {Nature}\ ,\ \bibinfo {pages} {1}} (\bibinfo {year} {2023})}\BibitemShut {NoStop}%
\bibitem [{\citenamefont {Xu}\ \emph {et~al.}(2023{\natexlab{a}})\citenamefont {Xu}, \citenamefont {Sun}, \citenamefont {Jia}, \citenamefont {Liu}, \citenamefont {Xu}, \citenamefont {Li}, \citenamefont {Gu}, \citenamefont {Watanabe}, \citenamefont {Taniguchi}, \citenamefont {Tong}, \citenamefont {Jia}, \citenamefont {Shi}, \citenamefont {Jiang}, \citenamefont {Zhang}, \citenamefont {Liu},\ and\ \citenamefont {Li}}]{xu2023observation}%
  \BibitemOpen
  \bibfield  {author} {\bibinfo {author} {\bibfnamefont {F.}~\bibnamefont {Xu}}, \bibinfo {author} {\bibfnamefont {Z.}~\bibnamefont {Sun}}, \bibinfo {author} {\bibfnamefont {T.}~\bibnamefont {Jia}}, \bibinfo {author} {\bibfnamefont {C.}~\bibnamefont {Liu}}, \bibinfo {author} {\bibfnamefont {C.}~\bibnamefont {Xu}}, \bibinfo {author} {\bibfnamefont {C.}~\bibnamefont {Li}}, \bibinfo {author} {\bibfnamefont {Y.}~\bibnamefont {Gu}}, \bibinfo {author} {\bibfnamefont {K.}~\bibnamefont {Watanabe}}, \bibinfo {author} {\bibfnamefont {T.}~\bibnamefont {Taniguchi}}, \bibinfo {author} {\bibfnamefont {B.}~\bibnamefont {Tong}}, \bibinfo {author} {\bibfnamefont {J.}~\bibnamefont {Jia}}, \bibinfo {author} {\bibfnamefont {Z.}~\bibnamefont {Shi}}, \bibinfo {author} {\bibfnamefont {S.}~\bibnamefont {Jiang}}, \bibinfo {author} {\bibfnamefont {Y.}~\bibnamefont {Zhang}}, \bibinfo {author} {\bibfnamefont {X.}~\bibnamefont {Liu}},\ and\ \bibinfo {author} {\bibfnamefont {T.}~\bibnamefont {Li}},\ }\bibfield  {title} {\bibinfo {title}
  {Observation of integer and fractional quantum anomalous hall effects in twisted bilayer ${\mathrm{mote}}_{2}$},\ }\href {https://doi.org/10.1103/PhysRevX.13.031037} {\bibfield  {journal} {\bibinfo  {journal} {Phys. Rev. X}\ }\textbf {\bibinfo {volume} {13}},\ \bibinfo {pages} {031037} (\bibinfo {year} {2023}{\natexlab{a}})}\BibitemShut {NoStop}%
\bibitem [{\citenamefont {Reddy}\ \emph {et~al.}(2023)\citenamefont {Reddy}, \citenamefont {Alsallom}, \citenamefont {Zhang}, \citenamefont {Devakul},\ and\ \citenamefont {Fu}}]{reddy2023fractional}%
  \BibitemOpen
  \bibfield  {author} {\bibinfo {author} {\bibfnamefont {A.~P.}\ \bibnamefont {Reddy}}, \bibinfo {author} {\bibfnamefont {F.}~\bibnamefont {Alsallom}}, \bibinfo {author} {\bibfnamefont {Y.}~\bibnamefont {Zhang}}, \bibinfo {author} {\bibfnamefont {T.}~\bibnamefont {Devakul}},\ and\ \bibinfo {author} {\bibfnamefont {L.}~\bibnamefont {Fu}},\ }\bibfield  {title} {\bibinfo {title} {Fractional quantum anomalous hall states in twisted bilayer ${\mathrm{mote}}_{2}$ and ${\mathrm{wse}}_{2}$},\ }\href {https://doi.org/10.1103/PhysRevB.108.085117} {\bibfield  {journal} {\bibinfo  {journal} {Phys. Rev. B}\ }\textbf {\bibinfo {volume} {108}},\ \bibinfo {pages} {085117} (\bibinfo {year} {2023})}\BibitemShut {NoStop}%
\bibitem [{\citenamefont {Wang}\ \emph {et~al.}(2024)\citenamefont {Wang}, \citenamefont {Zhang}, \citenamefont {Liu}, \citenamefont {He}, \citenamefont {Xu}, \citenamefont {Ran}, \citenamefont {Cao},\ and\ \citenamefont {Xiao}}]{wang2023fractional}%
  \BibitemOpen
  \bibfield  {author} {\bibinfo {author} {\bibfnamefont {C.}~\bibnamefont {Wang}}, \bibinfo {author} {\bibfnamefont {X.-W.}\ \bibnamefont {Zhang}}, \bibinfo {author} {\bibfnamefont {X.}~\bibnamefont {Liu}}, \bibinfo {author} {\bibfnamefont {Y.}~\bibnamefont {He}}, \bibinfo {author} {\bibfnamefont {X.}~\bibnamefont {Xu}}, \bibinfo {author} {\bibfnamefont {Y.}~\bibnamefont {Ran}}, \bibinfo {author} {\bibfnamefont {T.}~\bibnamefont {Cao}},\ and\ \bibinfo {author} {\bibfnamefont {D.}~\bibnamefont {Xiao}},\ }\bibfield  {title} {\bibinfo {title} {Fractional chern insulator in twisted bilayer ${\mathrm{mote}}_{2}$},\ }\href {https://doi.org/10.1103/PhysRevLett.132.036501} {\bibfield  {journal} {\bibinfo  {journal} {Phys. Rev. Lett.}\ }\textbf {\bibinfo {volume} {132}},\ \bibinfo {pages} {036501} (\bibinfo {year} {2024})}\BibitemShut {NoStop}%
\bibitem [{\citenamefont {Goldman}\ \emph {et~al.}(2023)\citenamefont {Goldman}, \citenamefont {Reddy}, \citenamefont {Paul},\ and\ \citenamefont {Fu}}]{goldman2023zero}%
  \BibitemOpen
  \bibfield  {author} {\bibinfo {author} {\bibfnamefont {H.}~\bibnamefont {Goldman}}, \bibinfo {author} {\bibfnamefont {A.~P.}\ \bibnamefont {Reddy}}, \bibinfo {author} {\bibfnamefont {N.}~\bibnamefont {Paul}},\ and\ \bibinfo {author} {\bibfnamefont {L.}~\bibnamefont {Fu}},\ }\bibfield  {title} {\bibinfo {title} {Zero-field composite fermi liquid in twisted semiconductor bilayers},\ }\href {https://doi.org/10.1103/PhysRevLett.131.136501} {\bibfield  {journal} {\bibinfo  {journal} {Phys. Rev. Lett.}\ }\textbf {\bibinfo {volume} {131}},\ \bibinfo {pages} {136501} (\bibinfo {year} {2023})}\BibitemShut {NoStop}%
\bibitem [{\citenamefont {Dong}\ \emph {et~al.}(2023)\citenamefont {Dong}, \citenamefont {Wang}, \citenamefont {Ledwith}, \citenamefont {Vishwanath},\ and\ \citenamefont {Parker}}]{dong2023composite}%
  \BibitemOpen
  \bibfield  {author} {\bibinfo {author} {\bibfnamefont {J.}~\bibnamefont {Dong}}, \bibinfo {author} {\bibfnamefont {J.}~\bibnamefont {Wang}}, \bibinfo {author} {\bibfnamefont {P.~J.}\ \bibnamefont {Ledwith}}, \bibinfo {author} {\bibfnamefont {A.}~\bibnamefont {Vishwanath}},\ and\ \bibinfo {author} {\bibfnamefont {D.~E.}\ \bibnamefont {Parker}},\ }\bibfield  {title} {\bibinfo {title} {Composite fermi liquid at zero magnetic field in twisted ${\mathrm{mote}}_{2}$},\ }\href {https://doi.org/10.1103/PhysRevLett.131.136502} {\bibfield  {journal} {\bibinfo  {journal} {Phys. Rev. Lett.}\ }\textbf {\bibinfo {volume} {131}},\ \bibinfo {pages} {136502} (\bibinfo {year} {2023})}\BibitemShut {NoStop}%
\bibitem [{\citenamefont {Morales-Dur\'an}\ \emph {et~al.}(2024)\citenamefont {Morales-Dur\'an}, \citenamefont {Wei}, \citenamefont {Shi},\ and\ \citenamefont {MacDonald}}]{morales2023magic}%
  \BibitemOpen
  \bibfield  {author} {\bibinfo {author} {\bibfnamefont {N.}~\bibnamefont {Morales-Dur\'an}}, \bibinfo {author} {\bibfnamefont {N.}~\bibnamefont {Wei}}, \bibinfo {author} {\bibfnamefont {J.}~\bibnamefont {Shi}},\ and\ \bibinfo {author} {\bibfnamefont {A.~H.}\ \bibnamefont {MacDonald}},\ }\bibfield  {title} {\bibinfo {title} {Magic angles and fractional chern insulators in twisted homobilayer transition metal dichalcogenides},\ }\href {https://doi.org/10.1103/PhysRevLett.132.096602} {\bibfield  {journal} {\bibinfo  {journal} {Phys. Rev. Lett.}\ }\textbf {\bibinfo {volume} {132}},\ \bibinfo {pages} {096602} (\bibinfo {year} {2024})}\BibitemShut {NoStop}%
\bibitem [{\citenamefont {Cr{\'e}pel}\ \emph {et~al.}(2023)\citenamefont {Cr{\'e}pel}, \citenamefont {Regnault},\ and\ \citenamefont {Queiroz}}]{crepel2023chiral}%
  \BibitemOpen
  \bibfield  {author} {\bibinfo {author} {\bibfnamefont {V.}~\bibnamefont {Cr{\'e}pel}}, \bibinfo {author} {\bibfnamefont {N.}~\bibnamefont {Regnault}},\ and\ \bibinfo {author} {\bibfnamefont {R.}~\bibnamefont {Queiroz}},\ }\bibfield  {title} {\bibinfo {title} {The chiral limits of moiré semiconductors: origin of flat bands and topology in twisted transition metal dichalcogenides homobilayers},\ }\href@noop {} {\bibfield  {journal} {\bibinfo  {journal} {arXiv preprint arXiv:2305.10477}\ } (\bibinfo {year} {2023})}\BibitemShut {NoStop}%
\bibitem [{\citenamefont {Reddy}\ and\ \citenamefont {Fu}(2023)}]{reddy2023toward}%
  \BibitemOpen
  \bibfield  {author} {\bibinfo {author} {\bibfnamefont {A.~P.}\ \bibnamefont {Reddy}}\ and\ \bibinfo {author} {\bibfnamefont {L.}~\bibnamefont {Fu}},\ }\bibfield  {title} {\bibinfo {title} {Toward a global phase diagram of the fractional quantum anomalous hall effect},\ }\href {https://doi.org/10.1103/PhysRevB.108.245159} {\bibfield  {journal} {\bibinfo  {journal} {Phys. Rev. B}\ }\textbf {\bibinfo {volume} {108}},\ \bibinfo {pages} {245159} (\bibinfo {year} {2023})}\BibitemShut {NoStop}%
\bibitem [{\citenamefont {Yu}\ \emph {et~al.}(2024)\citenamefont {Yu}, \citenamefont {Herzog-Arbeitman}, \citenamefont {Wang}, \citenamefont {Vafek}, \citenamefont {Bernevig},\ and\ \citenamefont {Regnault}}]{Yu2023fractional}%
  \BibitemOpen
  \bibfield  {author} {\bibinfo {author} {\bibfnamefont {J.}~\bibnamefont {Yu}}, \bibinfo {author} {\bibfnamefont {J.}~\bibnamefont {Herzog-Arbeitman}}, \bibinfo {author} {\bibfnamefont {M.}~\bibnamefont {Wang}}, \bibinfo {author} {\bibfnamefont {O.}~\bibnamefont {Vafek}}, \bibinfo {author} {\bibfnamefont {B.~A.}\ \bibnamefont {Bernevig}},\ and\ \bibinfo {author} {\bibfnamefont {N.}~\bibnamefont {Regnault}},\ }\bibfield  {title} {\bibinfo {title} {Fractional chern insulators versus nonmagnetic states in twisted bilayer ${\mathrm{mote}}_{2}$},\ }\href {https://doi.org/10.1103/PhysRevB.109.045147} {\bibfield  {journal} {\bibinfo  {journal} {Phys. Rev. B}\ }\textbf {\bibinfo {volume} {109}},\ \bibinfo {pages} {045147} (\bibinfo {year} {2024})}\BibitemShut {NoStop}%
\bibitem [{\citenamefont {Abouelkomsan}\ \emph {et~al.}(2024)\citenamefont {Abouelkomsan}, \citenamefont {Reddy}, \citenamefont {Fu},\ and\ \citenamefont {Bergholtz}}]{abouelkomsan2024band}%
  \BibitemOpen
  \bibfield  {author} {\bibinfo {author} {\bibfnamefont {A.}~\bibnamefont {Abouelkomsan}}, \bibinfo {author} {\bibfnamefont {A.~P.}\ \bibnamefont {Reddy}}, \bibinfo {author} {\bibfnamefont {L.}~\bibnamefont {Fu}},\ and\ \bibinfo {author} {\bibfnamefont {E.~J.}\ \bibnamefont {Bergholtz}},\ }\bibfield  {title} {\bibinfo {title} {Band mixing in the quantum anomalous hall regime of twisted semiconductor bilayers},\ }\href@noop {} {\bibfield  {journal} {\bibinfo  {journal} {Physical Review B}\ }\textbf {\bibinfo {volume} {109}},\ \bibinfo {pages} {L121107} (\bibinfo {year} {2024})}\BibitemShut {NoStop}%
\bibitem [{\citenamefont {Shi}\ \emph {et~al.}(2024)\citenamefont {Shi}, \citenamefont {Morales-Durán}, \citenamefont {Khalaf},\ and\ \citenamefont {MacDonald}}]{shi2024adiabatic}%
  \BibitemOpen
  \bibfield  {author} {\bibinfo {author} {\bibfnamefont {J.}~\bibnamefont {Shi}}, \bibinfo {author} {\bibfnamefont {N.}~\bibnamefont {Morales-Durán}}, \bibinfo {author} {\bibfnamefont {E.}~\bibnamefont {Khalaf}},\ and\ \bibinfo {author} {\bibfnamefont {A.~H.}\ \bibnamefont {MacDonald}},\ }\bibfield  {title} {\bibinfo {title} {Adiabatic approximation and aharonov-casher bands in twisted homobilayer transition metal dichalcogenides},\ }\bibfield  {journal} {\bibinfo  {journal} {Physical Review B}\ }\textbf {\bibinfo {volume} {110}},\ \href {https://doi.org/10.1103/physrevb.110.035130} {10.1103/physrevb.110.035130} (\bibinfo {year} {2024})\BibitemShut {NoStop}%
\bibitem [{\citenamefont {Li}\ and\ \citenamefont {Wu}(2024)}]{li2024variational}%
  \BibitemOpen
  \bibfield  {author} {\bibinfo {author} {\bibfnamefont {B.}~\bibnamefont {Li}}\ and\ \bibinfo {author} {\bibfnamefont {F.}~\bibnamefont {Wu}},\ }\href {https://arxiv.org/abs/2405.20307} {\bibinfo {title} {Variational mapping of chern bands to landau levels: Application to fractional chern insulators in twisted mote$_2$}} (\bibinfo {year} {2024}),\ \Eprint {https://arxiv.org/abs/2405.20307} {arXiv:2405.20307 [cond-mat.mes-hall]} \BibitemShut {NoStop}%
\bibitem [{\citenamefont {Morales-Dur{\'a}n}\ \emph {et~al.}(2023)\citenamefont {Morales-Dur{\'a}n}, \citenamefont {Wang}, \citenamefont {Schleder}, \citenamefont {Angeli}, \citenamefont {Zhu}, \citenamefont {Kaxiras}, \citenamefont {Repellin},\ and\ \citenamefont {Cano}}]{morales2023pressure}%
  \BibitemOpen
  \bibfield  {author} {\bibinfo {author} {\bibfnamefont {N.}~\bibnamefont {Morales-Dur{\'a}n}}, \bibinfo {author} {\bibfnamefont {J.}~\bibnamefont {Wang}}, \bibinfo {author} {\bibfnamefont {G.~R.}\ \bibnamefont {Schleder}}, \bibinfo {author} {\bibfnamefont {M.}~\bibnamefont {Angeli}}, \bibinfo {author} {\bibfnamefont {Z.}~\bibnamefont {Zhu}}, \bibinfo {author} {\bibfnamefont {E.}~\bibnamefont {Kaxiras}}, \bibinfo {author} {\bibfnamefont {C.}~\bibnamefont {Repellin}},\ and\ \bibinfo {author} {\bibfnamefont {J.}~\bibnamefont {Cano}},\ }\bibfield  {title} {\bibinfo {title} {Pressure-enhanced fractional chern insulators along a magic line in moir{\'e} transition metal dichalcogenides},\ }\href {https://doi.org/10.1103/PhysRevResearch.5.L032022} {\bibfield  {journal} {\bibinfo  {journal} {Physical Review Research}\ }\textbf {\bibinfo {volume} {5}},\ \bibinfo {pages} {L032022} (\bibinfo {year} {2023})}\BibitemShut {NoStop}%
\bibitem [{\citenamefont {Sharma}\ \emph {et~al.}(2024)\citenamefont {Sharma}, \citenamefont {Peng},\ and\ \citenamefont {Sheng}}]{sheng2024topological}%
  \BibitemOpen
  \bibfield  {author} {\bibinfo {author} {\bibfnamefont {P.}~\bibnamefont {Sharma}}, \bibinfo {author} {\bibfnamefont {Y.}~\bibnamefont {Peng}},\ and\ \bibinfo {author} {\bibfnamefont {D.~N.}\ \bibnamefont {Sheng}},\ }\bibfield  {title} {\bibinfo {title} {Topological quantum phase transitions driven by a displacement field in twisted ${\mathrm{mote}}_{2}$ bilayers},\ }\href {https://doi.org/10.1103/PhysRevB.110.125142} {\bibfield  {journal} {\bibinfo  {journal} {Phys. Rev. B}\ }\textbf {\bibinfo {volume} {110}},\ \bibinfo {pages} {125142} (\bibinfo {year} {2024})}\BibitemShut {NoStop}%
\bibitem [{\citenamefont {Lu}\ \emph {et~al.}(2024)\citenamefont {Lu}, \citenamefont {Wu}, \citenamefont {Chen}, \citenamefont {Sun},\ and\ \citenamefont {Meng}}]{lu2024interaction}%
  \BibitemOpen
  \bibfield  {author} {\bibinfo {author} {\bibfnamefont {H.}~\bibnamefont {Lu}}, \bibinfo {author} {\bibfnamefont {H.-Q.}\ \bibnamefont {Wu}}, \bibinfo {author} {\bibfnamefont {B.-B.}\ \bibnamefont {Chen}}, \bibinfo {author} {\bibfnamefont {K.}~\bibnamefont {Sun}},\ and\ \bibinfo {author} {\bibfnamefont {Z.~Y.}\ \bibnamefont {Meng}},\ }\href {https://arxiv.org/abs/2403.03258} {\bibinfo {title} {Interaction-driven roton condensation in c = 2/3 fractional quantum anomalous hall state}} (\bibinfo {year} {2024}),\ \Eprint {https://arxiv.org/abs/2403.03258} {arXiv:2403.03258 [cond-mat.str-el]} \BibitemShut {NoStop}%
\bibitem [{\citenamefont {Sheng}\ \emph {et~al.}(2024)\citenamefont {Sheng}, \citenamefont {Reddy}, \citenamefont {Abouelkomsan}, \citenamefont {Bergholtz},\ and\ \citenamefont {Fu}}]{sheng2024quantum}%
  \BibitemOpen
  \bibfield  {author} {\bibinfo {author} {\bibfnamefont {D.~N.}\ \bibnamefont {Sheng}}, \bibinfo {author} {\bibfnamefont {A.~P.}\ \bibnamefont {Reddy}}, \bibinfo {author} {\bibfnamefont {A.}~\bibnamefont {Abouelkomsan}}, \bibinfo {author} {\bibfnamefont {E.~J.}\ \bibnamefont {Bergholtz}},\ and\ \bibinfo {author} {\bibfnamefont {L.}~\bibnamefont {Fu}},\ }\bibfield  {title} {\bibinfo {title} {Quantum anomalous hall crystal at fractional filling of moir\'e superlattices},\ }\href {https://doi.org/10.1103/PhysRevLett.133.066601} {\bibfield  {journal} {\bibinfo  {journal} {Phys. Rev. Lett.}\ }\textbf {\bibinfo {volume} {133}},\ \bibinfo {pages} {066601} (\bibinfo {year} {2024})}\BibitemShut {NoStop}%
\bibitem [{\citenamefont {Xu}\ \emph {et~al.}(2023{\natexlab{b}})\citenamefont {Xu}, \citenamefont {Li}, \citenamefont {Xu}, \citenamefont {Bi},\ and\ \citenamefont {Zhang}}]{xu2023maximally}%
  \BibitemOpen
  \bibfield  {author} {\bibinfo {author} {\bibfnamefont {C.}~\bibnamefont {Xu}}, \bibinfo {author} {\bibfnamefont {J.}~\bibnamefont {Li}}, \bibinfo {author} {\bibfnamefont {Y.}~\bibnamefont {Xu}}, \bibinfo {author} {\bibfnamefont {Z.}~\bibnamefont {Bi}},\ and\ \bibinfo {author} {\bibfnamefont {Y.}~\bibnamefont {Zhang}},\ }\bibfield  {title} {\bibinfo {title} {Maximally localized wannier orbitals, interaction models and fractional quantum anomalous hall effect in twisted bilayer mote2},\ }\href {https://arxiv.org/abs/2308.09697} {\bibfield  {journal} {\bibinfo  {journal} {arXiv preprint arXiv:2308.09697}\ } (\bibinfo {year} {2023}{\natexlab{b}})}\BibitemShut {NoStop}%
\bibitem [{\citenamefont {Girvin}\ \emph {et~al.}(1986)\citenamefont {Girvin}, \citenamefont {MacDonald},\ and\ \citenamefont {Platzman}}]{girvin1986magneto}%
  \BibitemOpen
  \bibfield  {author} {\bibinfo {author} {\bibfnamefont {S.~M.}\ \bibnamefont {Girvin}}, \bibinfo {author} {\bibfnamefont {A.~H.}\ \bibnamefont {MacDonald}},\ and\ \bibinfo {author} {\bibfnamefont {P.~M.}\ \bibnamefont {Platzman}},\ }\bibfield  {title} {\bibinfo {title} {Magneto-roton theory of collective excitations in the fractional quantum hall effect},\ }\href {https://doi.org/10.1103/PhysRevB.33.2481} {\bibfield  {journal} {\bibinfo  {journal} {Phys. Rev. B}\ }\textbf {\bibinfo {volume} {33}},\ \bibinfo {pages} {2481} (\bibinfo {year} {1986})}\BibitemShut {NoStop}%
\bibitem [{\citenamefont {Wolf}\ \emph {et~al.}(2024)\citenamefont {Wolf}, \citenamefont {Chao}, \citenamefont {MacDonald},\ and\ \citenamefont {Su}}]{wolf2024intraband}%
  \BibitemOpen
  \bibfield  {author} {\bibinfo {author} {\bibfnamefont {T.~M.~R.}\ \bibnamefont {Wolf}}, \bibinfo {author} {\bibfnamefont {Y.-C.}\ \bibnamefont {Chao}}, \bibinfo {author} {\bibfnamefont {A.~H.}\ \bibnamefont {MacDonald}},\ and\ \bibinfo {author} {\bibfnamefont {J.~J.}\ \bibnamefont {Su}},\ }\href {https://arxiv.org/abs/2406.10709} {\bibinfo {title} {Intraband collective excitations in fractional chern insulators are dark}} (\bibinfo {year} {2024}),\ \Eprint {https://arxiv.org/abs/2406.10709} {arXiv:2406.10709 [cond-mat.str-el]} \BibitemShut {NoStop}%
\bibitem [{\citenamefont {Mao}\ \emph {et~al.}(2024)\citenamefont {Mao}, \citenamefont {Mendez-Valderrama},\ and\ \citenamefont {Chowdhury}}]{mao2024low}%
  \BibitemOpen
  \bibfield  {author} {\bibinfo {author} {\bibfnamefont {D.}~\bibnamefont {Mao}}, \bibinfo {author} {\bibfnamefont {J.~F.}\ \bibnamefont {Mendez-Valderrama}},\ and\ \bibinfo {author} {\bibfnamefont {D.}~\bibnamefont {Chowdhury}},\ }\href {https://arxiv.org/abs/2410.16352} {\bibinfo {title} {Is the low-energy optical absorption in correlated insulators controlled by quantum geometry?}} (\bibinfo {year} {2024}),\ \Eprint {https://arxiv.org/abs/2410.16352} {arXiv:2410.16352 [cond-mat.str-el]} \BibitemShut {NoStop}%
\bibitem [{\citenamefont {Regan}\ \emph {et~al.}(2020)\citenamefont {Regan}, \citenamefont {Wang}, \citenamefont {Jin}, \citenamefont {Bakti~Utama}, \citenamefont {Gao}, \citenamefont {Wei}, \citenamefont {Zhao}, \citenamefont {Zhao}, \citenamefont {Zhang}, \citenamefont {Yumigeta} \emph {et~al.}}]{regan2020mott}%
  \BibitemOpen
  \bibfield  {author} {\bibinfo {author} {\bibfnamefont {E.~C.}\ \bibnamefont {Regan}}, \bibinfo {author} {\bibfnamefont {D.}~\bibnamefont {Wang}}, \bibinfo {author} {\bibfnamefont {C.}~\bibnamefont {Jin}}, \bibinfo {author} {\bibfnamefont {M.~I.}\ \bibnamefont {Bakti~Utama}}, \bibinfo {author} {\bibfnamefont {B.}~\bibnamefont {Gao}}, \bibinfo {author} {\bibfnamefont {X.}~\bibnamefont {Wei}}, \bibinfo {author} {\bibfnamefont {S.}~\bibnamefont {Zhao}}, \bibinfo {author} {\bibfnamefont {W.}~\bibnamefont {Zhao}}, \bibinfo {author} {\bibfnamefont {Z.}~\bibnamefont {Zhang}}, \bibinfo {author} {\bibfnamefont {K.}~\bibnamefont {Yumigeta}}, \emph {et~al.},\ }\bibfield  {title} {\bibinfo {title} {Mott and generalized wigner crystal states in wse2/ws2 moir{\'e} superlattices},\ }\href {https://doi.org/10.1038/s41586-020-2092-4} {\bibfield  {journal} {\bibinfo  {journal} {Nature}\ }\textbf {\bibinfo {volume} {579}},\ \bibinfo {pages} {359} (\bibinfo {year} {2020})}\BibitemShut {NoStop}%
\end{thebibliography}%

\include{figures}

\setcounter{equation}{0}
\setcounter{figure}{0}
\setcounter{table}{0}
\setcounter{page}{1}
\makeatletter
\renewcommand{\theequation}{S\arabic{equation}}
\renewcommand{\thefigure}{S\arabic{figure}}
\renewcommand{\bibnumfmt}[1]{[S#1]}
\renewcommand{\citenumfont}[1]{S#1}
\renewcommand{\citenumfont}[1]{\textit{#1}}
\begin{widetext}

\begin{center}
{\huge \textbf{Supplemental Material}}
\end{center}
\quad \\\vskip 0.5cm

\begin{center}
\textbf{ I. STRUCTURE FACTOR FORMALISM}
\end{center}

\begin{center}
\textbf{A. Explicit calculation}
\end{center}

The static structure factor is defined in the main text is the Fourier
transform of the static density-density correlation function, $\Tilde{S}(\bm{q})=\frac{1}{A}\langle\rho(\bm{q})\rho(\bm{-q})\rangle$ for general fermionic momentum-space density operators $\rho(\bm{q})=\sum_{\bm{k}}f^\dagger_{\bm{k}}f_{\bm{k}+\bm{q}}$ in the plane wave basis. Here in the supplement, we use the standard convention employed in exact diagonalization studies: $S(\bm{q})=\frac{A}{N_e}\Tilde{S}(\bm{q}) = \frac{1}{N_e}\langle\rho(\bm{q})\rho(\bm{-q})\rangle$ for number of electrons $N_e$ and system area $A$. This convention used in the supplement is to make $S(\bm{q})$ dimensionless, which eases notation of the calculations made in the supplement. The convention used in the main text is done to make quantum weight $K$ dimensionless, as it had been defined in previous works \cite{onishi2024quantum}. Referring to equation 7 in the main text, combining the definition of the fermionic momentum-space density operators in the continuum model with the definition of the structure factor, the explicit form of the static structure factor becomes

    \begin{multline}
    S(\bm{q}) =\frac{1}{N_e}\left\langle\sum_{\sigma_1,X_1,\bm{k_1}} f^\dagger_{\sigma_1,X_1,\bm{k_1}}f_{\sigma_1,X_1,\bm{k_1}+\bm{q}} \sum_{\sigma_2,X_2,\bm{k_2}} f^\dagger_{\sigma_2,X_2,\bm{k_2}}f_{\sigma_2,X_2,\bm{k_2}-\bm{q}}\right\rangle \\
     = \frac{1}{N_e} \sum_{\substack{\sigma_1,X_1,\bm{k_1}\\\sigma_2,X_2,\bm{k_2}}} \Bigg\langle \sum_{\substack{n_1,n_2,n_3,n_4\\\bm{G},\bm{G'},\bm{G''},\bm{G'''} }}u^*_{\sigma_1,n_1;\bm{G},X_1}(\bm{k_1}) c^\dagger_{\sigma_1,n_1,\bm{k_1}} u_{\sigma_1,n_2;\bm{G'},X_1}(\bm{k_1}+\bm{q})c_{\sigma_1,n_2,\bm{k_1}+\bm{q}}\\ \times  u^*_{\sigma_2,n_3;\bm{G''},X_2}(\bm{k_2}) c^\dagger_{\sigma_2,n_3,\bm{k_2}} u_{\sigma_2,n_4;\bm{G'''},X_2}(\bm{k_2}-\bm{q})c_{\sigma_2,n_4,\bm{k_2}-\bm{q}}\Bigg\rangle.
\end{multline}

Using standard manipulation of fermionic raising and lowering operator algebra and conservation of momentum, we arrive at the full expression for the structure factor: 

\begin{multline}
    S(\bm{q})=\frac{1}{N_e}\sum_{\substack{\sigma_1,\sigma_2,X_1,\\X_2,\bm{k_1},\bm{k_2}}} \sum_{\substack{n_1,n_2,\\n_3,n_4}}\sum_{\substack{\bm{G},\bm{G'},\\ \bm{G''},\bm{G'''}}} \Bigg[ \delta_{\sigma_1,\sigma_2}\delta_{n_2,n_3}\delta_{\bm{k_2},\bm{k_1}+\bm{q}}  u^*_{\sigma_1,n_1;\bm{G},X_1}(\bm{k_1}) u^*_{\sigma_2,n_3;\bm{G''},X_2}(\bm{k_2}) \\ \times u_{\sigma_2,n_4;\bm{G'},X_2}(\bm{k_2}-\bm{q}) u_{\sigma_1,n_2;\bm{G'''},X_1}(\bm{k_1}+\bm{q}) \left \langle  c^\dagger_{\sigma_1,n_1,\bm{k_1}} c_{\sigma_2,n_4,\bm{k_2}-\bm{q}} \right \rangle \\
    + u^*_{\sigma_1,n_1;\bm{G},X_1}(\bm{k_1}) u^*_{\sigma_2,n_3;\bm{G'},X_2}(\bm{k_2}) u_{\sigma_2,n_4;\bm{G''},X_2}(\bm{k_2}-\bm{q}) u_{\sigma_1,n_2;\bm{G'''},X_1}(\bm{k_1}+\bm{q}) \\ \times \left\langle c^\dagger_{\sigma_1,n_1,\bm{k_1}} c^\dagger_{\sigma_2,n_3,\bm{k_2}} c_{\sigma_2,n_4,\bm{k_2}-\bm{q}} c_{\sigma_1,n_2,\bm{k_1}+\bm{q}} \right \rangle \Bigg ],
\end{multline}

which simplifies to our final analytic expression:

\begin{multline}
    S(\bm{q})=\frac{1}{N_e} \sum_{\substack{\sigma_1,X_1,\\X_2,\bm{k_1}}} \sum_{\substack{n_1,n_2,n_3 \\ \bm{G},\bm{G'},\bm{G''}, \bm{G'''}}} \Bigg[ u^*_{\sigma_1,n_1;\bm{G},X_1}(\bm{k_1}) u^*_{\sigma_1,n_3;\bm{G''},X_2}(\bm{k_1}+\bm{q}) u_{\sigma_1,n_2;\bm{G'},X_2}(\bm{k_1}) u_{\sigma_1,n_3;\bm{G'''},X_1}(\bm{k_1}+\bm{q}) \\ \times \left \langle  c^\dagger_{\sigma_1,n_1,\bm{k_1}} c_{\sigma_1,n_2,\bm{k_1}} \right \rangle \\
    +  \sum_{\sigma_2, \bm{k_2}, n_4} u^*_{\sigma_1,n_1;\bm{G},X_1}(\bm{k_1}) u^*_{\sigma_2,n_3;\bm{G''},X_2}(\bm{k_2}) u_{\sigma_2,n_4;\bm{G'},X_2}(\bm{k_2}-\bm{q}) u_{\sigma_1,n_2;\bm{G'''},X_1}(\bm{k_1}+\bm{q}) \\ \times \left\langle c^\dagger_{\sigma_1,n_1,\bm{k_1}} c^\dagger_{\sigma_2,n_3,\bm{k_2}} c_{\sigma_2,n_4,\bm{k_2}-\bm{q}} c_{\sigma_1,n_2,\bm{k_1}+\bm{q}} \right \rangle \Bigg ]. \label{final_Sq}
\end{multline}

Going forward, the first term will be referred to as the single body component of the structure factor (Since it only contains the momentum of one particle), $S_1(\bm{q})$, and the second term will be referred to as the normal ordered two body component (where $\left\langle : \rho(\bm{q})\rho(\bm{-q}) : \right\rangle \equiv \left\langle c^\dagger c^\dagger c c\right\rangle$ denotes normal ordering), $S_2(\bm{q})$, such that $S(\bm{q})=\frac{1}{N_e}\left[S_1(\bm{q})+ S_2(\bm{q})\right]$.
\par \hfill \break

\begin{center}
\textbf{B. Numerical calculation}
\end{center}

In practice, calculating the second-quantized fermionic operators for all bands is too computationally expensive, which makes using the explicit form of the full structure factor cumbersome in numerical calculations. Remarkably, we can calculate the full structure factor solely from the components of the projected structure factor. To see this, let us first represent the structure factor in position space and decompose it into our one and two body components (assuming identical atoms):

\begin{equation}
\begin{split}
    S(\bm{q}) & =\frac{1}{N_e}\left\langle\sum_{i,j}e^{-i\bm{q}(\bm{r}_i-\bm{r}_j)}\right\rangle \\ & = \frac{1}{N_e}\left\langle\sum_{i=j}e^{-i\bm{q}(\bm{r}_i-\bm{r}_j)}+\sum_{i\neq j}e^{-i\bm{q}(\bm{r}_i-\bm{r}_j)}\right\rangle \\ & =\frac{1}{N_e}\left[S_1(\bm{q})+ S_2(\bm{q})\right]  =  1+ \frac{1}{N_e}S_2(\bm{q}).
\end{split}
\end{equation}

Now we show that assuming the many-body ground state lives in the projected space, $P|\Psi_{GS}\rangle=|\Psi_{GS}\rangle$ for projection operator $P$, then 

\begin{align}
\begin{split}
    \Bar{S_2}(\bm{q})&=\left\langle:\Bar{\rho}(\bm{q})\Bar{\rho}(\bm{-q}):\right\rangle \\ & = \left\langle P \sum_{i\neq j}e^{-i\bm{q}\bm{r}_i}P^2 \sum_{j\neq i}e^{-i\bm{q}\bm{r}_j}P\right\rangle \\ &= \left\langle P \sum_{i\neq j}e^{-i\bm{q}(\bm{r}_i-\bm{r}_j)}P^3\right\rangle \\ & = \left\langle\sum_{i\neq j}e^{-i\bm{q}(\bm{r}_i-\bm{r}_j)}\right\rangle = S_2(\bm{q}).
\end{split} \label{S_2ofq}
\end{align}

Finally, since the projected structure factor can also be factorized into one and two body components, $\Bar{S}(\bm{q})=\frac{1}{N_e}\left[\Bar{S_1}(\bm{q})+ \Bar{S_2}(\bm{q})\right]$, the full structure factor takes on the following form, exclusively in terms of the projected components: 

\begin{equation}
    S(\bm{q})=1 + \frac{1}{N_e}\Bar{S_2}(\bm{q}) = 1+ \left[\Bar{S}(\bm{q}) - \frac{1}{N_e}\Bar{S_1}(\bm{q})\right]. \label{SofqToSq_2}
\end{equation}

The band projected structure factor is simply:

\begin{equation}
    \Bar{S}(\bm{\Tilde{q}})=\frac{1}{N_e}\left\langle \Bar{\rho}(\bm{\Tilde{q}})\Bar{\rho}(\bm{-\Tilde{q}})\right\rangle_{\bm{GS}}
\end{equation}

where $\Bar{\rho}(\bm{\Tilde{q}})=\sum_{\bm{\Tilde{k}}}\Bar{f}^\dagger_{\bm{\Tilde{k}}}\Bar{f}_{\bm{\Tilde{k}}+\bm{\Tilde{q}}}$ is the band projected  fermionic momentum-space density
operators , $N_e$ is the number of electrons in the system, $\bm{\Tilde{k}}$, $\bm{\Tilde{q}}$ are vectors in the MBZ, and $\left\langle \right\rangle_{\bm{GS}}$ represents averaging over all degenerate ground states. Note that for the ordinary structure factor, we would be dealing with $\bm{k}=\bm{\Tilde{k}}+\bm{g}$ and $\bm{q}=\bm{\Tilde{q}}+\bm{g}$ for $\bm{k}$ and $\bm{q}$ in the Brillouin Zone (BZ), and $\bm{g}$ a reciprocal lattice vector. WLOG and to ease notation, we denote $\bm{k}$ and $\bm{q}$ as momentum vectors in the MBZ for the remainder of this section. Thus, we work out an explicit form for our band-projected structure factor:

\begin{multline}\label{proStruct}
    \Bar{S}(\bm{q})=\frac{1}{N_e}\left\langle \sum_{\sigma_1,\bm{k_1}}\Bar{f}^\dagger_{\sigma_1,\bm{k_1}}\Bar{f}_{\sigma_1,\bm{k_1}+\bm{q}} \sum_{\sigma_2,\bm{k_2}}\Bar{f}^\dagger_{\sigma_2,\bm{k_2}}\Bar{f}_{\sigma_2,\bm{k_2}-\bm{q}} \right\rangle_{\bm{GS}}\\
    =\frac{1}{N_e} \sum_{\substack{\sigma_1, \bm{g},\bm{k_1} \\ n_1, n_2}} \Bigg[|F(\bm{k_1},\bm{k_1},n_1,n_2,\sigma_1,\bm{g}_{\bm{q}})|^2\left\langle c^\dagger_{\sigma_1,n_1,\bm{k_1}} c_{\sigma_1,n_2,\bm{k_1}}\right\rangle_{\bm{GS}} \\ + \sum_{\substack{\sigma_2,\bm{g'},\bm{k_2},\\ n_3,n_4}}F(\bm{k_1},\bm{k_1}+\bm{q},n_1,n_2,\sigma_1,\bm{g}_{\bm{q}}) F(\bm{k_2},\bm{k_2}-\bm{q},n_3,n_4,\sigma_2,\bm{g'}_{-\bm{q}}) \left\langle c^\dagger_{\sigma_1,n_1,\bm{k_1}} c^\dagger_{\sigma_2,n_3,\bm{k_2}} c_{\sigma_2,n_4,\bm{k_2}-\bm{q}} c_{\sigma_1,n_2,\bm{k_1}+\bm{q}} \right \rangle _{\bm{GS}}\Bigg] \\ =  \frac{1}{N_e}\left[\Bar{S_1}(\bm{q}) + \Bar{S_2}(\bm{q})\right], 
\end{multline}

where $\{n_i\}_{i=1,2,3,4}$ run over the projected bands, $\bm{g}_{\bm{q}}\equiv \bm{q}-[\bm{q}]$ is the reciprocal lattice component of an arbitrary momentum vector $\bm{q}$, and $F(\bm{k_1},\bm{k_2},n_1, n_2, \sigma,\bm{g})=\sum_{\bm{g'},X}u_{\sigma, n_1;\bm{(\bm{g'}+\bm{g})},X}^{*}(\bm{k_1})u_{\sigma,n_2;\bm{\bm{g'},X}}(\bm{k_2})$ is the form factor coming from the normalization of single particle Bloch states labeled by moiré crystal momentum, band index and spin, $|\bm{k},n,\sigma\rangle = \sum_{\bm{g},X}u_{\sigma,n;\bm{g},X}(\bm{k})|\bm{k}+\bm{g},n,X,\sigma\rangle$. The operator $\left[ \right]$ replaces an arbitrary momentum vector with its ``mesh" equivalent $\bm{k}$, where each $\bm{k}$ is an unique representative of crystal momentum within the MBZ ``mesh". From this convention, it follows that any arbitrary momentum vector, $\bm{q}=[\bm{q}]+\bm{g}_{\bm{q}}$, can be uniquely represented on a computational grid. Thus, the explicit form for $S(\bm{q})$ is simply $1 + \frac{1}{N_e}\Bar{S}_2(\bm{q})$ where $\Bar{S}_2(\bm{q})$ is the second term in equation \ref{proStruct}.
\par \hfill \break

\begin{center}
\textbf{II. FIT EXTRACTED QUANTUM WEIGHT}
\end{center}

In this section, we provide a brief overview for how the quantum weight can be extracted from the full structure factor. As mentioned in the main text, the quantum weight, $K$, is the leading order coefficient of the structure factor in materials with $U(1)$ symmetry:

\bea
    \Tilde{S}(\bm{q})= \frac{K_{\mu \nu}}{2\pi}q_\mu q_\nu + \cdots,
\eea

where the sum over spatial indices is implied \cite{onishi2024StructureFact,onishi2024quantum}. It can also be shown by through a quantum geometric formulation of the structure factor $\Tilde{S}(\bm{q)}=\int_{BZ} \frac{d^d \bm{k}}{(2\pi)^d} \text{Tr}\left[P(\bm{k})\big(P(\bm{k})-P(\bm{k}+\bm{q}\big)\right]$ for projection operator $P(\bm{k})$, that $K$ is simply the integral of the Fubini-Study metric of the occupied bands over the BZ: $K = 2\pi \int_{BZ} \frac{d^d \bm{k}}{(2\pi)^d} \sum_\mu g_{\mu \mu}$. However, band theory fails for interacting systems, and the quantum weight must be formulated in terms of the many body ground state quantum geometric tensor with twisted boundary conditions:

\begin{equation}
    K = 2\pi \int \frac{d^d \bm{\kappa}}{(2\pi)^d} \sum_{\mu}Tr\left(Re(\mathcal{Q_{\mu \mu}})\right),
\end{equation}

for $\mathcal{Q}_{\mu\nu}=\langle \frac{\partial}{\partial \kappa_\mu}\Psi_{\bm{\kappa}}|(1-P_{\bm{\kappa}})|\frac{\partial}{\partial \kappa_\nu}\Psi_{\bm{\kappa}}\rangle$, projection operator onto the ground state subspace $P_{\bm{\kappa}}$, and $\bm{\kappa}$ specifies the twisted boundary conditions over the ground state $\Psi_{\bm{\kappa}}(\bm{r_1},...,\bm{r_i}+\bm{L_\mu},...,\bm{r_{N_e}})=\Psi_{\bm{\kappa}}(\bm{r_1},...,\bm{r_i},...,\bm{r_{N_e}})$, with $\bm{L_\mu}=(0,...,L_\mu,...,0)$ specifying the system size in $\mu$ direction \cite{onishi2023fundamental}. In practice, calculating the many body quantum geometric tensor $\mathcal{Q}_{\mu\nu}$ for interacting systems is both analytically and numerically challenging, but we can calculate $K$ through the structure factor, by fitting the low $\bm{q}$ behavior and extracting the quadratic coefficient. Since $S(\bm{q})$ is an even function, we only need to use even-powered expansion coefficients in our fit. Further $S(\bm{q}=0)=0$, making the first point in our fit fully determined which requires the inclusion of at least the next 2 points to avoid over-fitting with a single quadratic fit parameter. Since the next two points in the 27 site cluster along the $\gamma$ to $\kappa$ linecut are 1/4 and 3/4 of the way to $\kappa$ from $\gamma$ respectively, fitting the next two points with only 1 quadratic coefficient has high variance and does not truly capture the $q\rightarrow 0$ behavior. As a result, we fit the first 4 points ($\gamma$ to $\kappa$), using $q^2$ and $q^4$ coefficients and extract the $q^2$ coefficient as the quantum weight $K$ (up to a numerical factor of $\frac{4\pi A}{N_e}=2\sqrt{3}\pi\frac{a_M^2}{\nu}$). The variance in the fitting is provided by the error bars in Fig. 5 in the main text. In Fig \ref{figS1: K vs N}, we show that this method of extracting $K$ converges nicely in the non-interacting case where we can calculate $K$ for larger system sizes. Fig. \ref{figS1: K vs N} demonstrates the minimal finite size effects in the quantum weight extraction for the 27 site cluster. The 27 site cluster (pictured in the inset of Fig. 2 in main text) has a favorable geometry with a maximal point density in the $\gamma$ to $\kappa$ line-cut, which in effect limits the finite size impact on the extracted quantum weight. Indeed the extracted quantum weight for the 27 site cluster is closer to the convergent value of $K=1.069$ than the next 5 system sizes (up to $N=121$), demonstrating the robustness of the extracted quantum weight on the 27 site cluster.

\begin{figure*}[t]
\centering
        \includegraphics[width=0.6\linewidth]{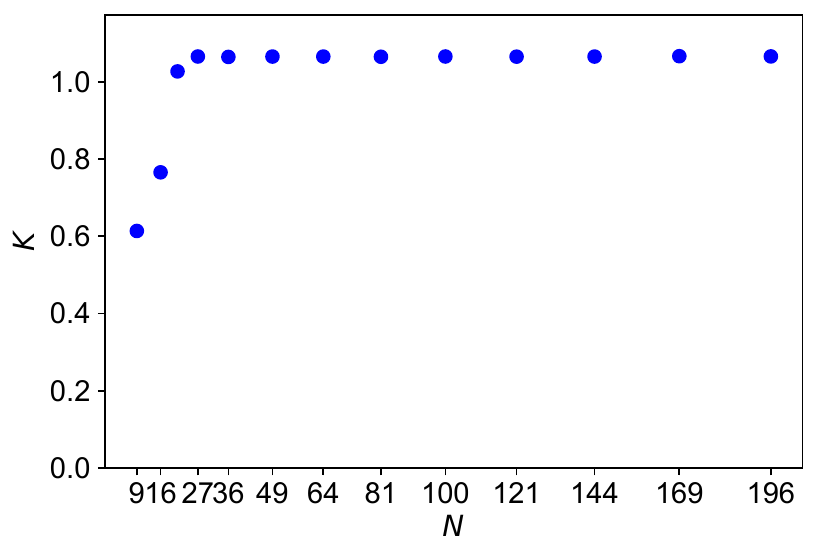} \quad \\\vskip 0.5cm
    \caption{Extracted quantum weight as a function of system size for $\nu=1$ at $\theta=1.75^\circ$, $\epsilon=10$, and $D=0$. }
    \label{figS1: K vs N}
\end{figure*}

\par \hfill \break

\begin{center}
\textbf{III. BRAGG PEAK SCALING}
\end{center}

In order to evaluate how Bragg peaks scale with system size, we first note the expression for $S(\bm{q})$ with the Bragg peak divergence subtraced, otherwise known as the disconnected part of $S(\bm{q})$: 
\begin{equation}
    S(\bm{q})_{disconnected}=\frac{1}{N_e}\big(\langle\rho(\bm{q})\rho(\bm{-q})\rangle- \langle\rho(\bm{q})\rangle \langle\rho(\bm{-q})\rangle \big)
\end{equation}

\begin{figure*}[t]
\centering
\begin{minipage}{0.99\textwidth}
        \includegraphics[width=0.8\linewidth]{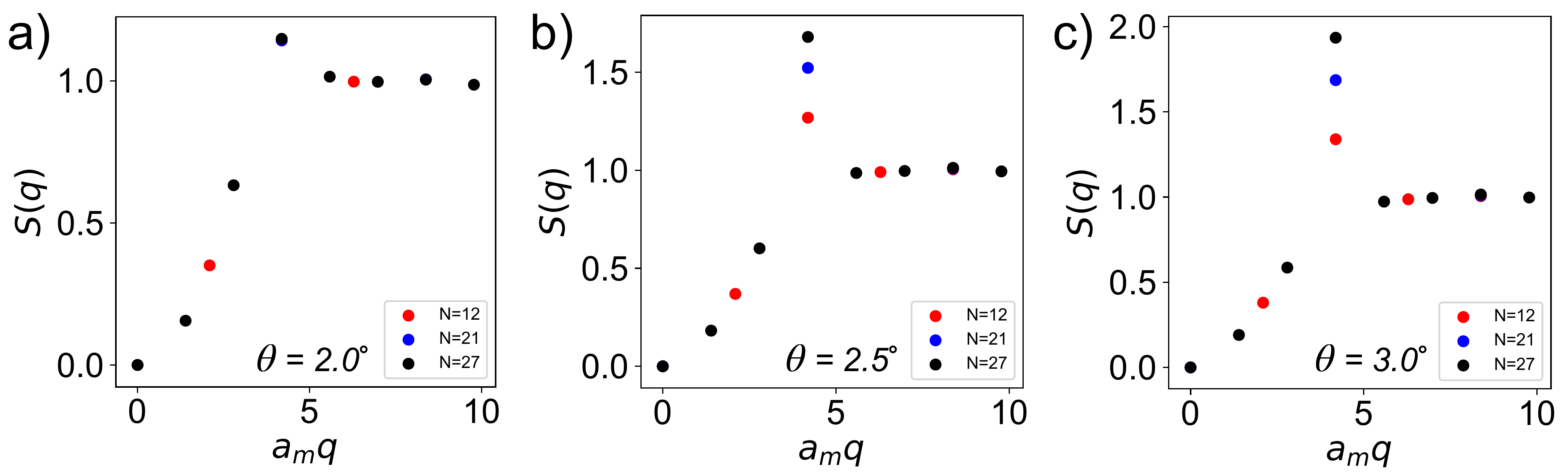} \quad \\\vskip 0.5cm
    \caption{$S(\bm{q})$ vs $q$ along $\gamma$ to $\kappa$ linecut for $\nu=1/3$ at different system sizes. Panels (a-c) are demonstrated for $\theta=2.0^\circ,2.5^\circ,3.0^\circ$ respectively.}
    \label{figS2: Bragg linecut}
\end{minipage}
\end{figure*}

\begin{figure*}[t]
\centering
\begin{minipage}{0.99\textwidth}
        \includegraphics[width=0.8\linewidth]{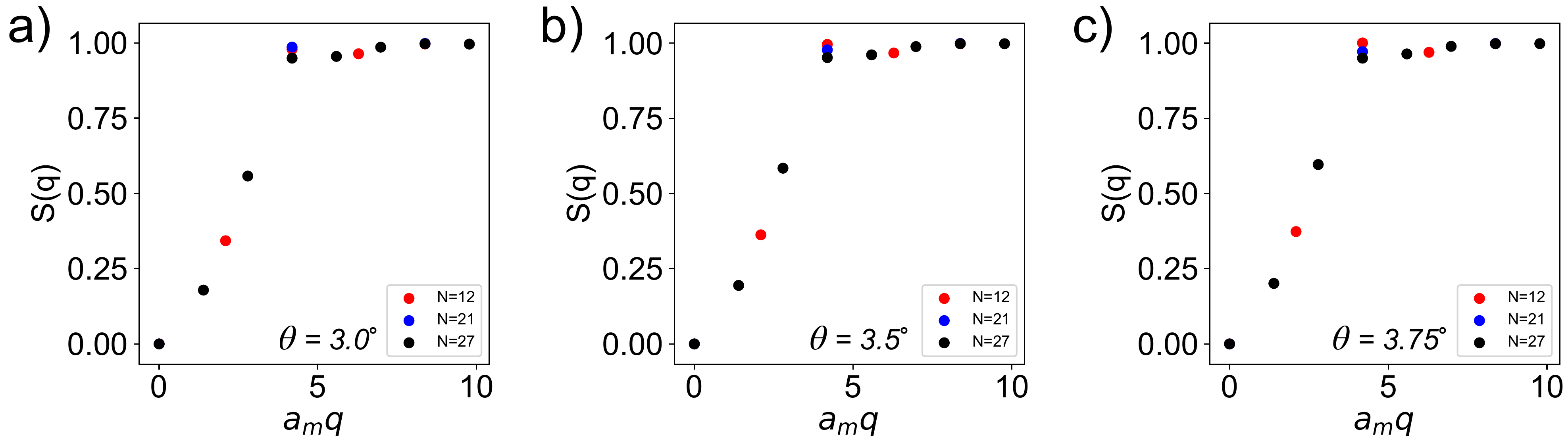} \quad \\\vskip 0.5cm
    \caption{$S(\bm{q})$ vs $q$ along $\gamma$ to $\kappa$ linecut for $\nu=2/3$ at different system sizes. Panels (a-c) are demonstrated for $\theta=3.0^\circ,3.5^\circ,3.75^\circ$ respectively.}
    \label{figS3: Bragg linecut 2}
\end{minipage}
\end{figure*}

The Bragg peaks themselves will scale as $S(\bm{q}) - S(\bm{q})_{disconnected} = \frac{1}{N_e}\langle\rho(\bm{q})\rangle \langle\rho(\bm{-q})\rangle = \frac{1}{N_e}n(\bm{k})$, which is just the orbital occupation number divided by the particle number. Thus since the density operator scales on the order of system size, the Bragg peaks will also scale linearly with system size. In Fig. \ref{figS2: Bragg linecut}, $S(\bm{q})$ is plotted for $\nu=1/3$ at different system sizes for increasing $\theta$. As $\theta$ changes from $2^\circ$ to $3^\circ$, $t$MoTe$_2$ transitions from an FCI to a GWC, just like we show in the main text at $\theta=2.0^\circ$ as displacement field increases. While this transition at $\nu=1/3$ is already well known \cite{reddy2023fractional}, we use the clear FCI to GWC transition to demonstrate the finite scaling signature of a crystal. With the current parameters, the system transitions from a FCI to a GWC at $\theta \approx 2.3^\circ$; however, the phases compete near this angle. From Fig. \ref{figS2: Bragg linecut}, peaks at $\kappa, \kappa'$ clearly emerge as twist angle increases, and the Bragg peak scaling becomes conspicuously linear at $\theta=2.5^\circ$, conclusively demonstrating the Bragg peak signature of the electron crystal, as mentioned in the main text. 

In sharp contrast, Fig.  \ref{figS3: Bragg linecut 2} shows that for $\theta>3.0^\circ$, where the system is outside the known FCI range, no clear crystal signature emerges. While Bragg peaks are present, they do not scale linearly with system size, indicating a non-pure crystalline state. These results add clarity to the potential phases beyond $\theta=3.0^\circ$ and reinforce the power fo the structure factor to help diagnosis phases of strongly correlated systems.
\par \hfill \break

\begin{center}
\textbf{IV. INTERACTION ENERGY AND QUANTUM WEIGHT}
\end{center}

As mentioned in the main text, we can relate the interaction energy, $\langle V \rangle_{GS}$, to the structure factor $S(\bm{q})$ since they both can be expressed in terms of fermionic momentum-space density operators. Specifically, 

\begin{equation}
    E_{int} = \langle\bm{V}\rangle = \left\langle\frac{1}{2} \sum_{\substack{\sigma,\sigma'; \bm{k'_1}\\ n_1,n_2,n_3,n_4;\\ \bm{k'_2}\bm{k_1}, \bm{k_2}}} V_{\bm{k'_1}\bm{k'_2}\bm{k_1}\bm{k_2}}^{\sigma\sigma'} c^\dagger_{\sigma_1,n_1,\bm{k'_1}} c^\dagger_{\sigma_2,n_3,\bm{k'_2}} c_{\sigma_2,n_4,\bm{k_2}} c_{\sigma_1,n_2,\bm{k_1}}\right\rangle, \label{E_int}
\end{equation}

\begin{figure*}[t]
\centering

    \includegraphics[width=0.8\linewidth]{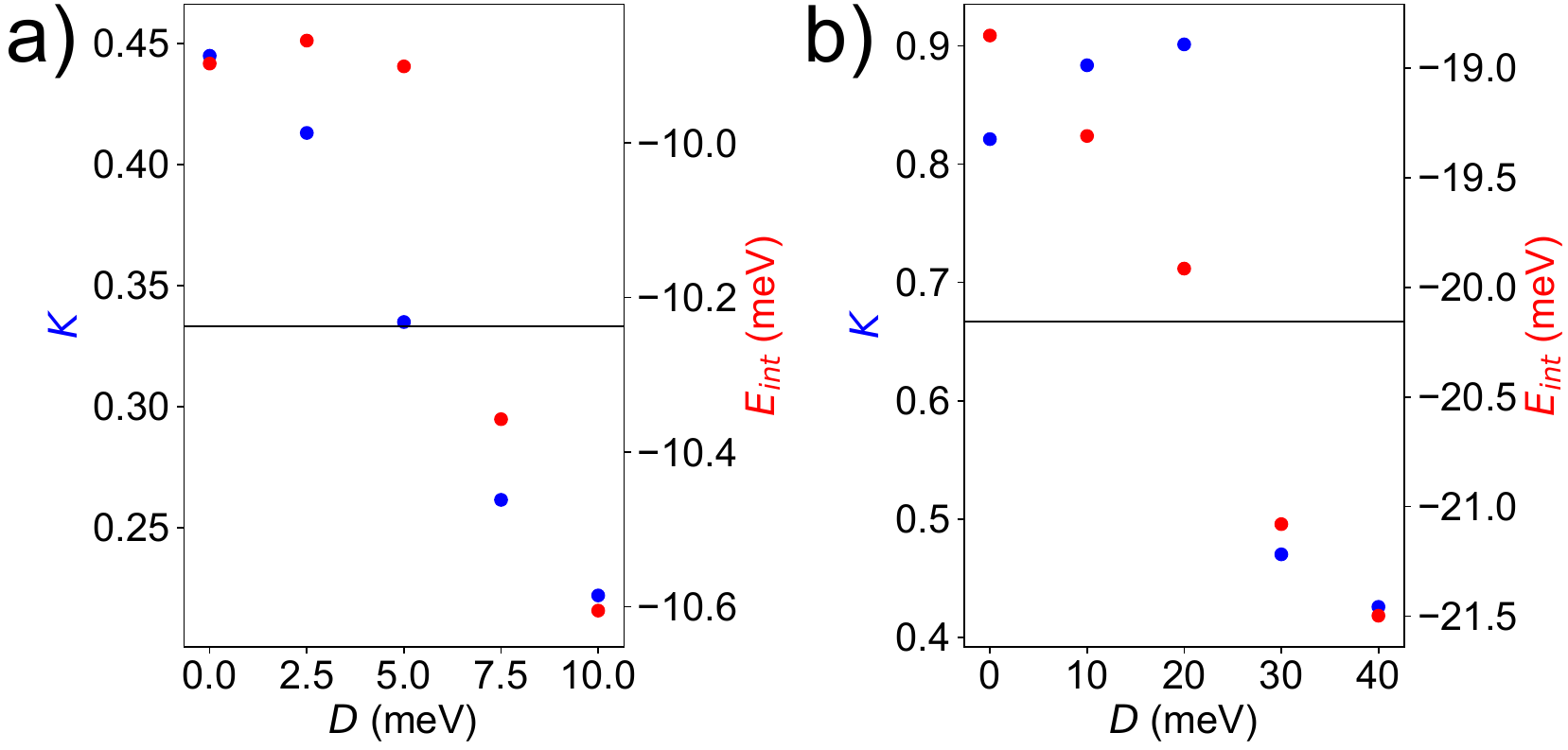} \quad \\\vskip 0.5cm
    \caption{(a,b) Quantum weight and Interaction Energy as a function of displacement field for $\nu=1/3,2/3$ respectively. The drop in interaction energy is accompanied by drop in quantum weight, explaining why quantum weight experiences a sharp decrease in the GWC phase.}
    \label{figS4: E_int}
\end{figure*}

while Eq. \ref{final_Sq} shows the similar form for $S(\bm{q})$. If we equate Eq. \ref{E_int} with the second term in \ref{final_Sq} (the normal ordered term) and sum over all $\bm{q}$, we arrive at the result:

\bea
E_{\rm int} = \sum_{\bm{q}}\frac{V(\bm{q}) \left(S_2(\bm{q})/N_e\right)}{2A} = \sum_{\bm{q}}\frac{V(\bm{q}) \left( S(\bm{q})-1\right)}{2A}
\eea

where Coulomb interaction $V(\bm{q}) = 2\pi e^2/\epsilon |\bm{q}|$ is positive at all $\bm{q}$, $N_e$ is the number of electrons in the system, and $A$ is system area, as stated in the main text (The last equality follows from Eq. \ref{SofqToSq_2}). In practice, we calculate $\Bar{S}_2(\bm{q})$, which implies $E_{\rm int} = \sum_{\bm{q}}\frac{V(\bm{q}) \left(\Bar{S}_2(\bm{q})/N_e\right)}{2A}$ by Eq. \ref{S_2ofq}. As noted in section I., the use of the band-projected structure factor is valid when the ground state wavefunction lives in the projected band subspace, which is indeed true for all systems studied in this paper. We noted that interaction energy decreases over the FCI-to-GWC transition since electrons localize to a configuration that minimizes energy. However the emergence of Bragg peaks indicates a spike of $S(\bm{q})$ at the reciprocal lattice harmonic wavevector. Thus, if interaction energy is lowered over this transition, $S(\bm{q})$ must decrease considerably for all other $\bm{q}$, namely for $\bm{q}\rightarrow 0$. We demonstrate this phenomena numerically in Fig. \ref{figS4: E_int} by plotting the interaction energy and quantum weight side by side as the system undergoes the FCI-to-GWC transition under increasing displacement field. 

\par \hfill \break

\begin{center}
\textbf{V. EXTENDED DATA}
\end{center}

We present an identical figure to Fig. 5 in the main text at $\epsilon=10$ rather than $\epsilon=5$ for completeness. For $\nu=2/3$, there is a level crossing in the ED spectrum around $\theta=3.0^\circ$, demonstrating evidence for a phase transition, although the nature of this phase is not precisely known \cite{reddy2023fractional}. Apart from this unknown phase region, there are no considerable qualitative differences between the Fig. 5 and \ref{figS5: K vs N}, with the quantum weight slightly lower for higher twist angles and slightly below rather than above the topological lower bound for $\nu=1/3$ at $\theta=2.5^\circ$.

\begin{figure*}[t]
\centering
    \includegraphics[width=0.55\linewidth]{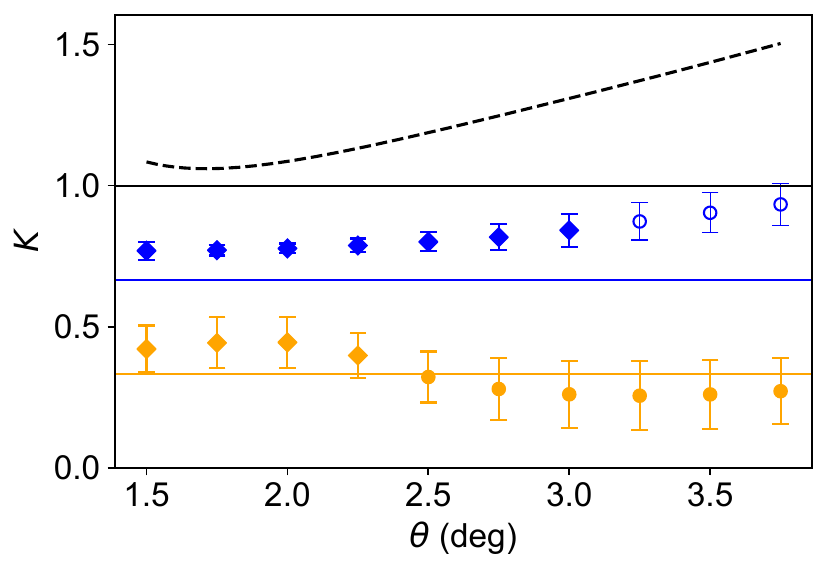} \quad \\\vskip 0.5cm
    \caption{Illustration of the universal topological bound being respected at fractional fillings. Extracted quantum weight is plotted versus twist angle at $\nu=1/3,2/3,1$ (orange, blue, black respectively). The universal topological bound for each filling is plotted as a solid horizontal line in the appropriate color, and the phase of the system at each angle is denoted by the style of the point. The diamonds correspond to points within the FCI phase; the solid circles correspond to points within the GWC phase; the hallow circles correspond to unconfirmed phases \cite{reddy2023fractional}. The dashed black line is the non-interacting case, previously shown to respect the topological bound in this system at integer filling \cite{onishi2024StructureFact}. All data is calculated at $\epsilon=10$ and $D=0$ on a 27 site cluster (error bars reflect the fit uncertainty due to finite size effects).}
    \label{figS5: K vs N}
\end{figure*}

\end{widetext}

\end{document}